\documentstyle[12pt]{article}

\advance\hoffset by -7mm  
\makeatletter
\@addtoreset{equation}{section}
\makeatother

\renewcommand{\title}[1]{\null\vspace{25mm}

\noindent{\Large{\bf #1}}\vspace{10mm}

\noindent {\large By }}
\newcommand{\authors}[1]{\noindent{\large #1}\vspace{3mm}

}
\newcommand{\address}[1]{\noindent #1\vspace{5mm}

}
\renewcommand{\abstract}[1]{\vspace{19mm}

\noindent{\small{\em Abstract.} #1}\vspace{2mm}

} 

\newcommand{\be}{\begin{equation}}
\newcommand{\ee}{\end{equation}}
\newcommand{\bea}{\begin{eqnarray}}
\newcommand{\eea}{\end{eqnarray}}
\newcommand{\bdm}{\begin{displaymath}}
\newcommand{\edm}{\end{displaymath}}

\newcommand{\cod}{d^{\dagger}}
\newcommand{\we}{\wedge}

\newcommand{\gtens}{\mbox{\boldmath $g$}}
\newcommand{\deltens}{\mbox{\boldmath $\delta$}}
\newcommand{\gfour}{^{(4)} \gtens}
\newcommand{\Gtens}{\mbox{\boldmath $G$}}
\newcommand{\gatens}{\mbox{\boldmath $\gamma$}}
\newcommand{\sigtens}{\mbox{\boldmath $\sigma$}}
\newcommand{\Ttens}{\mbox{\boldmath $T$}}

\newcommand{\gtiltens}{\tilde{\gtens}}
\newcommand{\gtil}{\tilde{g}}

\newcommand{\ghattens}{\hat{\gtens}}

\newcommand{\Sgfold}{(\Sigma,\gtens)}
\newcommand{\nabdel}{\underline{\nabla}}
\newcommand{\nabtil}{\tilde{\nabla}}
\newcommand{\laptil}{\tilde{\Delta}}

\newcommand{\cotil}{\tilde{d}^{\dagger}}

\newcommand{\sprod}[2]{\langle #1 \, , #2 \rangle}

\newcommand{\commut}[2]{[ #1 \, , #2 ]}
\newcommand{\trace}[1]{\mbox{tr} \left\{ #1 \right\}}
\newcommand{\Trace}[1]{\mbox{Tr} \left\{ #1 \right\}}
\newcommand{\tfree}[1]{\stackrel{\circ}{#1}}
\newcommand{\idid}{1 \! \! 1}
\newcommand{\Txi}{T(\xi)}

\newcommand{\Erpot}{\mbox{$\rm E$}}

\begin{document}

\title{No-Hair Theorems and Black Holes with Hair}
\authors{Markus Heusler
}
\address{Institute for Theoretical Physics \\
The University of Zurich \\
CH--8057 Zurich, Switzerland}

\abstract{The critical steps leading to the uniqueness theorem 
for the Kerr-Newman metric are examined in the light of the new 
black hole solutions with Yang-Mills and scalar hair. Various 
methods -- including scaling techniques, arguments based on 
energy conditions, conformal transformations and divergence 
identities -- are reviewed, and their range of application to 
selfgravitating scalar and non-Abelian gauge fields is discussed. 
In particular, the no-hair theorem is extended to harmonic mappings 
with arbitrary Riemannian target manifolds. This paper is an extended
version of an invited lecture held at the 
{\em Journ\'ees Relativistes 96}.}

\section{Introduction}

The uniqueness theorem for the asymptotically
flat, stationary black hole solutions of the Einstein-Maxwell
equations is, by now, established quite rigorously 
(see, e.g. \cite{PC-94-DG} and \cite{MH-96-LN}).
Some open gaps, notably the electrovac staticity theorem 
\cite{SW-92-93} and the topology theorem (see
\cite{G-96-CQG} and the following
lecture by P. Chru\'sciel \cite{PTC-lect}) 
have been closed recently.
The theorem, conjectured by Israel, Penrose and Wheeler
in the late sixties (see \cite{WI-87-300} for a historical account), 
implies that all stationary electrovac black hole spacetimes are 
characterized by their mass, angular momentum and electric charge.
This beautiful result -- together with the striking analogy between
the laws of black hole physics and the laws of equilibrium 
thermodynamics -- provided support for the expectation that {\em all\/} 
stationary black hole solutions can be described in terms of a 
small set of asymptotically measurable quantities.

Tempting as it was, the hypothesis was disproved in 1989, when
several authors presented a counterexample within the framework 
of SU(2) Einstein-Yang-Mills (EYM) theory \cite{VG-89-JETP}:
Although the new solution was static and had vanishing Yang-Mills 
charges, it was different from the Schwarzschild black hole and,
therefore, {\em not\/} characterized by its total mass. 
In fact, a whole variety of new black hole configurations violating 
the generalized no-hair conjecture were found during the last few
years. These include, for instance, black holes with Skyrme 
\cite{DHS-91-PLB}, dilaton \cite{LM-93-NPB} or Yang-Mills-Higgs 
hair \cite{BFM-92-NPB}.
The diversity of new solutions gives rise to a reexamination of the 
logic to the proof of the uniqueness theorem.
In particular, one needs to investigate whether there are steps in 
the uniqueness proof which are not sensitive to the details of the 
matter contents.
In order to do so, we shall now briefly recall some of
the main issues which are involved in the uniqueness program.

\section{The Uniqueness Program}

At the basis of the reasoning lies Hawking's {\em strong rigidity\/} 
theorem \cite{H-72-CMP}, \cite{HE-73}.
It relates the concept of the event horizon to the independently 
defined -- and logically distinct -- local notion of the Killing 
horizon:
Requiring that the fundamental matter fields obey well behaved 
hyperbolic equations and that the stress-energy tensor satisfies the 
weak energy condition, the theorem asserts that the event horizon of 
a {\em stationary\/} black hole spacetime is a Killing horizon. 
This also implies that either the null-generator Killing field of the
horizon coincides 
with the stationary Killing field or spacetime admits at least
one axial Killing field. 
(The original proof of the rigidity theorem was based 
on an analyticity assumption which has, for instance, no justification
if the domain of outer communications admits regions where the 
stationary Killing field becomes null or spacelike; see the lecture by 
P. Chru\'sciel.)
The strong rigidity theorem implies that stationary black hole 
spacetimes are either axisymmetric or have a nonrotating horizon.
The classical uniqueness theorems were, however, established for spacetimes 
which are either {\em circular\/} or {\em static.\/}
Hence, in both cases, one has to prove in advance that
the Frobenius {\em integrability conditions\/} for the Killing fields
are satisfied as a consequence of the 
symmetry properties and the matter equations.

The {\em circularity\/} theorem, due to Kundt and Tr\"umper \cite{KT-66} 
and Carter \cite{C-69}, \cite{C-87}, implies that the 
metric of a vacuum or electrovac spacetime can, without loss of
generality, be written in the well-known Papapetrou  
$(2 \! + \! 2)$-split. As we shall show in section 5,
the generalization of the
theorem is straightforward for selfgravitating scalar 
mappings \cite{MH-93}
(i.e., Higgs fields, harmonic mappings, Skyrme fields, etc.).
The circularity theorem does, however, {\em not\/} 
hold for the EYM system
(without imposing additional constraints on the gauge fields
by hand). 

The {\em staticity\/} theorem, establishing the hypersurface 
orthogonality of the stationary Killing field for electrovac
black hole spacetimes with
nonrotating horizons, is more involved than the 
circularity problem.
First, one has to establish {\em strict\/} 
stationarity, that is, one needs to
exclude ergoregions. This problem, first discussed by Hajicek 
\cite{PH-73-75} and Hawking and Ellis \cite{HE-73}, 
was solved only recently by Sudarsky and Wald \cite{SW-92-93},
assuming a foliation by maximal slices \cite{PC-94-MS}. 
Once ergoregions are excluded, it remains to prove that the
stationary Killing field satisfies the
Frobenius integrability conditions. 
This was achieved by Hawking \cite{H-72-CMP}, extending a theorem 
due to Lich\-ne\-ro\-wicz \cite{L-55} to the {\em vacuum black hole\/} 
case. As already mentioned, Sudarsky and Wald were eventually 
able to solve the staticity problem for {\em electrovac\/} 
black holes by using the generalized version of the first law of 
black hole physics \cite{SW-92-93}.
Like the circularity theorem, the staticity theorem is easily 
extended to scalar fields \cite{MH-93}, whereas the problem is 
again open for selfgravitating non-Abelian gauge fields.

The main task of the uniqueness problem is to show that the static
electrovac black hole spacetimes (with nondegenerate horizon) are
described by the Reissner-Nordstr\"om metric, whereas the circular
ones (i.e., the stationary and axisymmetric ones with integrable 
Killing fields) are given by the Kerr-Newman metric.

In the static case it was Israel who, in his pioneering work
\cite{WI-67}, \cite{WI-68},
was able to establish that both static vacuum and electrovac
black hole spacetimes are {\em spherically symmetric\/}. 
Israel's ingenious method, based on integral identities and Stokes' 
theorem, triggered a series of papers devoted to the
uniqueness problem (see, e.g. \cite{MZH-73-74}, \cite{DCR77}). 
More recently, Bunting and Masood-ul-Alam \cite{BM-87}
(see also \cite{WS-85}) 
found a new proof of the Israel theorem, 
taking essential advantage of the positive mass theorem 
\cite{SY-79-81}, \cite{Wi-81}.

The uniqueness theorem for stationary and axisymmetric black holes
is mainly based on the Ernst formulation of the Einstein (-Maxwell)
equations \cite{Ernst-68}. The key result consists in Carter's 
observation that the field equations can be reduced to a 
$2$-dimensional boundary-value problem \cite{BC-71-73a-73c}.
A (most amazing) identity due to Robinson \cite{DCR-75} then establishes 
that all vacuum solutions with the same boundary and regularity conditions
are identical. The uniqueness problem for the electrovac case remained
open until Mazur \cite{Maz-82-84b} and Bunting \cite{Bunt-83} 
independently succeeded in deriving the desired divergence identities
in a systematic way: The Mazur identity is based on the observation 
that the Ernst equations describe a nonlinear sigma-model on the 
coset space $G/H$, where $G$ is a connected Lie group and $H$ is 
a maximal compact subgroup of $G$. In the electrovac case one finds 
$G/H = SU(1,2) / S(U(1) \times U(2))$. Within Mazur's approach, the 
Robinson identity turns out to be the explicit form of the sigma-model
identity for the vacuum case, $G/H = SU(1,1) / U(1)$.

\section{Selfgravitating Soliton Solutions}

One of the reasons why it was not until 1989 that black hole 
solutions with selfgravitating gauge fields were discovered was
the widespread belief that the EYM equations admit {\em no soliton\/} 
solutions either. There were, at least, four good reasons in support of 
this hypothesis.\\
$\bullet$ First, there are {\em no purely gravitational solitons\/}, that is, 
the only globally regular, asymptotically flat, static vacuum 
solution of Einstein's equations with finite energy is Minkowski 
spacetime.
Using Stokes' theorem for a spacelike hypersurface $\Sigma$,
this result is obtained from the positive mass theorem \cite{SY-79-81},
\cite{Wi-81} and the Komar expression for the total mass of an 
asymptotically flat, stationary spacetime with Killing field 
($1$-form) $k$, say:
\be
M \, = \, -\frac{1}{8 \pi G} \int_{S^{2}_{\infty}} \ast dk \, = \,
-\frac{1}{4 \pi G} \int_{\Sigma} \ast R(k) \, = \, 0 \, .
\label{Komarmass}
\ee
Here we have also used the vacuum Einstein equations and 
the Ricci identity $d \ast d k = 2 \ast R(k)$ 
(where the $1$-form $R(k)$ is defined by
$R(k)_{\mu} \equiv R_{\mu \nu} k^{\nu}$).\\ 
$\bullet$ Second, both Deser's energy argument \cite{SD-76} and Coleman's 
scaling method \cite{Col-77} show that there exist 
{\em no flat spacetime solitons in pure YM\/} theory.\\
$\bullet$ Third, the {\em Einstein-Maxwell\/} equations admit
{\em no soliton\/} solutions. 
This follows immediately from Stokes' theorem 
\be
\int_{\partial \Sigma} \ast (k \we \alpha) \, = \, 
-\int_{\Sigma} (\cod \alpha) \ast k
\label{Stokes}
\ee
for an arbitrary, invariant $1$-form $\alpha$ ($L_{k} \alpha = 0$),
and from the static Maxwell equations 
($V \equiv -\sprod{k}{k} \equiv - k^{\mu} k_{\mu}$) 
\be
\cod \left( \frac{E}{V} \right) \, = \, 0 \, , \; \; \; 
\cod \left( \frac{B}{V} \right) \, = \, 0 \, , \; \; \; 
E \, = \, d \phi \, , \; \; \; dB \, = \, 0 \, ,
\label{Maxwell}
\ee
for the electric ($E \equiv -i_{k}F$) and the 
magnetic ($B \equiv i_{k} \ast F$) component of the 
field strength $2$-form $F$ (see, e.g. \cite{MH-96-LN}).
(Here $\cod \equiv \ast d \ast$ denotes the coderivative operator and
$\ast k = i_{k} \eta$ is the volume $3$-form on $\Sigma$.)
First choosing $\alpha =  \frac{E}{V}$ gives
$-4 \pi Q \equiv \int_{\partial \Sigma} \ast (k \we \frac{E}{V}) = 0$. 
Then setting $\alpha =  \phi \frac{E}{V}$ yields 
$0 = -4 \pi Q \phi_{\infty} = \int_{\Sigma} \frac{\sprod{E}{E}}{V}$, 
and therefore $E=0$. 
A similar argument shows that the magnetic field vanishes as well.\\
$\bullet$ Finally, Deser \cite{SD-84} 
has shown that the {\em $3$-dimensional\/} 
EYM equations admit {\em no soliton solutions\/} 
either. His argument relies
on the fact that the magnetic $1$-form, $B = i_{k} \ast F$,
has only one nonvanishing component in $2+1$ dimensions.

All this shows that it was conceivable to conjecture a
nonexistence theorem for selfgravitating EYM solitons (in $3+1$ dimensions). 
On the other hand, none of the above examples takes care of
the full nonlinear structure of both the gauge and the gravitational fields. 
It is therefore, with hindsight, not too surprising that EYM solitons
actually do exist -- although it came, of course, unexpected
when Bartnik and McKinnon (BK) presented the first particle-like
solution in 1988 \cite{BK-88}. 
Since the BK solutions are spherically symmetric, it is instructive
to consider the implications of scaling methods for static,
spherically symmetric matter models coupled to gravity.

\section{Scaling Techniques}
 
Scaling arguments provide an efficient tool
for proving nonexistence theorems in flat spacetime.
Moreover, experience shows that
they also give reliable hints concerning the possible existence of
such solutions. In relativity, scaling techniques are,
of course, limited to highly symmetric situations, since they
are based on the existence of distinguished coordinates. In this
section we shall, therefore, restrict ourselves to 
{\em spherically symmetric\/},
selfgravitating field theories. (Here we present a slightly 
improved version of the arguments given in \cite{MHNS-92}.)

The effective action for static, spherically symmetric soliton
configurations is 
\be
{\cal A} \, [m,S,\psi] \, = \, \int_{0}^{\infty} 
\left( L_{\mbox{mat}} \, - \, S \, m' \right) \, dr \, ,
\label{action}
\ee
where $\psi$ stands for the matter fields with
effective matter Lagrangian $L_{\mbox{mat}}[m,S,\psi]$, and 
$m(r)$ and $S(r)$ parametrize the static spacetime metric,
\be
\gtens \, = \, - S^2 N \, dt^2 \, + \, N^{-1} \, dr^2 
\, + \, r^2 \, d \Omega^2 \, , \; \; \; \mbox{with} \; \;
N(r) \equiv 1 - \frac{2 \, m(r)}{r} \, .
\label{metric}
\ee
The complete set of static field equations is equivalent
to the Euler-Lagrange equations for $m$, $S$ and the matter fields.

Let us now assume that we are given a solution $m(r)$, $S(r)$,
$\psi(r)$ with boundary conditions $m(0) = m_0$, 
$m(\infty) = m_{\infty}$ etc. Then 
each member of the $1$-parameter family
\be
m_{\lambda}(r) \equiv m(\lambda r) \, , \; \; \; 
S_{\lambda}(r) \equiv S(\lambda r) \, , \; \; \; 
\psi_{\lambda}(r) \equiv \psi(\lambda r) 
\label{scaling}
\ee
assumes the same boundary values
at $r = 0$ and $r = \infty$, and the action
${\cal A}_{\lambda} 
\equiv {\cal A}[m_{\lambda}, S_{\lambda}, \psi_{\lambda}]$
must therefore have a critical point at $\lambda = 1$,
$\left[ d {\cal A} / d \lambda \right]_{\lambda = 1} = 0$.
Since the purely gravitational part does not
depend on $\lambda$ 
(since $\int_0^{\infty} dr \,
S_{\lambda}(r) [\partial \, m_{\lambda}(r) / \partial r]$ $=$
$\int_0^{\infty} d \rho \,
S(\rho) \, [\partial m(\rho) / \partial \rho]$, with 
$\rho = \lambda r$), we obtain the following {\em necessary} condition
for the existence of soliton solutions:
\be
\left[ \frac{d}{d \lambda} \int_{0}^{\infty} L_{\mbox{mat}}
\left[ m_{\lambda}(r), S_{\lambda}(r), \psi_{\lambda}(r) \right] \, dr
\right]_{\lambda = 1} \, = \, 0 \, .
\label{virial}
\ee
(It is worth noticing that gravity {\em does\/} enter the
argument, although the purely gravitational part
of the action does not appear in the above formula.)

As a first example, we consider the purely magnetic
SU(2) configuration used by Bartnik and McKinnon 
\cite{BK-88} (Witten ansatz) with gauge potential
$1$-form $A$,
\be
A \, = \, [w(r) -1] \, \left[
\tau_{\varphi} \, d \vartheta \, - \, 
\tau_{\vartheta} \sin \! \vartheta \, d \varphi \right] \, ,
\label{BMKansatz}
\ee
where $\tau_{\vartheta} \equiv \partial_{\vartheta} \tau_r$,
$\tau_{\varphi} \sin \! \vartheta \equiv \partial_{\varphi} \tau_r$
and $\tau_r \equiv (2i |\vec{r}|)^{-1} \vec{r} \cdot \vec{\sigma}$.
The effective matter Lagrangian then becomes (see, e.g. \cite{BHS-96})
\be
L_{\mbox{mat}} \, = \, S \, \left[ N T_{Y\!M} \, + \, V_{Y\!M} \right] \, ,
\; \; \; \mbox{where} \; \; \; 
T_{Y\!M} \equiv \frac{1}{2} w'^2 \, , \; \; \; 
V_{Y\!M} \equiv \frac{(1-w^2)^2}{4 \, r^2} \, .
\label{YMlagr}
\ee
A short computation shows that the condition 
(\ref{virial}) now implies that every solution must satisfy
\be
\int_{0}^{\infty} dr \, S  \, \left[ N T_{Y\!M} \, + \, 
V_{Y\!M} \right] \,  = \,
\int_{0}^{\infty} dr \, \frac{2 m}{r} \, S \, T_{Y\!M} \, .
\label{vir-YM}
\ee
In flat spacetime the same argument yields 
$\int dr \, [ T_{Y\!M} + V_{Y\!M} ] = 0$, which shows that
no solitons exist. If, however, gravity is taken into account,
then the virial type relation (\ref{vir-YM}) does 
{\em not exclude\/} selfgravitating soliton solutions any more.

A second, even simpler example are scalar 
(Higgs) fields with non-negative
potentials $P[\phi]$. In this case we find
\be
L_{\mbox{mat}} \, = S \, \left[ N T_H \, + \, V_H \right] \, ,
\; \; \; \mbox{where} \; \; \; 
T_H \equiv \frac{1}{2} r^2 \phi'^2 \, , \; \; \; 
V_H \equiv r^2 P[\phi] \, .
\label{Hlagr}
\ee
The scaling argument now yields the condition
$\int dr \, S \left[ T_H + 3 V_H \right] = 0$,
which is -- up to the positive function $S(r)$ -- the same relation 
as one obtains in flat spacetime. Hence, as is well known, there are 
no spherically symmetric scalar solitons 
(with non-negative potentials), independently of whether or not gravity
is taken into account.
In fact, the above technique can be applied to spherically
symmetric {\em harmonic\/} mappings with arbitrary 
Riemannian target manifolds (see also the next section).

The above scaling argument can also be used to exclude -- or 
indicate -- the existence of {\em black hole\/} solutions.
To this end, one has to replace the lower boundary in the
action integral (\ref{action}) by the horizon distance $r_H$.
In order to have fixed boundary conditions 
(at $r = r_H$ and $r = \infty$) we consider the
modified $1$-parameter family 
\be
m_{\lambda}(r) \equiv m(r_H + \lambda \cdot (r-r_H)) \, , \; \; \; 
S_{\lambda}(r) \equiv S(r_H + \lambda \cdot (r-r_H))
\label{scaling-2}
\ee 
(and similarly for the matter field amplitudes).
The same argument as before shows that every solution is
subject to the condition (\ref{virial}), where now the functions 
$m_{\lambda}$ etc. are given by eq. (\ref{scaling-2}) and the 
lower boundary in the integral is $r = r_H$.
Using the effective YM and scalar field Lagrangians (\ref{YMlagr}) 
and (\ref{Hlagr}) we obtain the virial relations
\be
\int_{r_H}^{\infty} dr \, S \, \left\{
[ 1 + \frac{2m}{r} \left( \frac{r_H}{r} - 2 \right) ] \, T_{Y\!M}
\, + \, 
[ 1 - \frac{2 r_H}{r} ] \, V_{Y\!M} \right\} \, = \, 0
\label{bh-YM}
\ee
and
\be
\int_{r_H}^{\infty} dr \, S \, \left\{
[ \frac{2 r_H}{r} \left( 1 - \frac{m}{r} \right) - 1 ] \, T_H
\, + \, 
[ \frac{2 r_H}{r} - 3 ] \, V_H \right\} \, = \, 0 \, ,
\label{bh-H}
\ee
respectively. 
It is obvious that the factors in front of the (non-negative)
YM quantities $T_{Y\!M}$ and $V_{Y\!M}$ in eq. (\ref{bh-YM}) are indefinite.
Hence, in the EYM case, the scaling argument does {\em not\/} exclude 
black hole solutions with hair. In contrast to this,
both factors in front of the (non-negative) quantities
$T_H$ and $V_H$ in eq. (\ref{bh-H}) {\em do\/} 
have a fixed, negative sign.
(Use $[\ldots](r_H) = 0$ and $[\ldots]'(r) \leq 0 \, \forall r \geq r_H$
(since $m' \geq 0$ as a consequence of the dominant energy condition)
to show that the bracket in front of the kinetic term $T_H$ is
nonpositive.)
Hence, the virial relation (\ref{bh-H}) is violated, unless
$T_H = V_H = 0$, which excludes spherically
symmetric black holes with nontrivial scalar fields. In fact,
(a slightly more complicated version of) this argument provided the
first proof of the no-hair theorem for spherically symmetric
scalar fields with {\em arbitrary\/} non-negative
potentials \cite{MHNS-92}.
In the meantime, Sudarsky \cite{DS-NH} and 
Bekenstein \cite{B-NH} have presented
different proofs of the spherically symmetric scalar no-hair theorem.
There exists, yet, another proof which is exclusively based
on a mass bound for spherically symmetric black holes and
the circumstance that scalar fields (with harmonic action and
non-negative potentials) 
violate the strong energy condition \cite{H-MB}.
The following section is devoted to a brief
outline of this argument.

\section{Uniqueness Theorems and Mass Bounds}

The total mass of a stationary, asymptotically flat spacetime
with Killing field $k$ is given by the Komar formula
(\ref{Komarmass}). Applying Stokes' theorem and the
Ricci identity for Killing fields, one obtains the Smarr formula
\be
M \, = \, \frac{1}{4 \pi} \kappa {\cal A} \, - \, \frac{1}{4 \pi}
\int_{\Sigma} \ast R(k) \, ,
\label{Smarr}
\ee
where $\kappa$ and ${\cal A}$ denote the surface gravity and the
area of the horizon, respectively, and where we have assumed that 
the latter is nonrotating (i.e., generated by $k$). Now using
the identity $d \omega = \ast (k \we R(k))$ for the derivative of 
the twist ($\omega \equiv \frac{1}{2} \ast [k \we dk]$), 
as well as the general identity (\ref{twist-id}) below, 
it is not hard to derive the formula
(see, e.g. \cite{MH-96-LN} or \cite{H-MB})
\be
M \, = \, \frac{1}{4 \pi} \kappa {\cal A} \, + \, \frac{1}{4 \pi}
\int_{\Sigma} \left[
\frac{R(k,k)}{V} \, - \, 2 \, \frac{\sprod{\omega}{\omega}}{V^2} \right]
\ast k \, ,
\label{Smarr2}
\ee
where, as before, $V \equiv - \sprod{k}{k} \equiv - k^{\mu} k_{\mu}$.
If the domain of outer communications is {\em strictly\/} stationary,
then $k$ is nowhere spacelike and $\omega$ is nowhere timelike. 
The violation of the {\em strong\/} energy condition (SEC) 
(by which we mean $R(k,k) \leq 0$ throughout the domain
of outer communications) then implies an 
{\em upper\/} bound for the total mass:
\be 
M \, \leq \; \frac{1}{4 \pi} \kappa {\cal A} \, , \;
\; \; \; \mbox{if} \; \; \; 
\Ttens(k,k) \, - \, \frac{1}{2} \, 
\gtens(k,k) \,\mbox{tr} \, \Ttens \, \leq \, 0 \, .
\label{ADM8} 
\ee 

If no horizons are present, this inequality reduces to $M \leq 0$,
which is in contradiction to the {\em positive mass\/} theorem 
\cite{SY-79-81}, \cite{Wi-81} -- unless $M = 0$.
Since the latter is based on the {\em dominant\/} energy
condition (DEC), we conclude that there exist 
{\em no soliton\/} solutions in 
selfgravitating field theories which are subject to the DEC, but
violate the SEC at every point.

For static {\em black hole\/} solutions of matter models which
violate the SEC we have $M \leq \frac{1}{(4 \pi)} \kappa {\cal A}$.
Hence, in order to extend the above soliton argument to the black hole
case, a derivation of the converse inequality, 
$M \geq \frac{1}{(4 \pi)} \kappa {\cal A}$,
on the basis of the DEC is needed.
However, this was not achieved until now.
In fact, the Penrose conjecture \cite{P-73},
$M^2 \geq {\cal A}/(16 \pi)$, which is closely related 
to the above inequality, was also
proven only under additional geometrical assumptions
\cite{LV-83} or in the spherically symmetric case
\cite{MM-MB}, \cite{SH-MB}
(see also \cite{BG-73}, \cite{JW-77} and \cite{H-MB}).
In the general case, the strongest result consists in the extension of
the positive mass theorem by Gibbons {\it{et al.}} \cite{G-PM} to
{\em black} {\em hole\/} spacetimes. It is, however, 
not hard to establish the desired bound, 
$M \geq \frac{1}{(4 \pi)} \kappa {\cal A}$, 
for spherically symmetric configurations (see below). 
This yields the conclusion that all static, spherically symmetric 
black hole solutions with matter
satisfying the DEC and violating the SEC
coincide with the Schwarzschild metric.

Before we apply this result to scalar fields, we owe the 
derivation of $M \geq \frac{1}{(4 \pi)} \kappa {\cal A}$ for
static, spherically symmetric black hole spacetimes with matter
being subject to the DEC. First, using the parametrization
(\ref{metric}) of the metric, the DEC implies that both 
$m(r)$ and $S(r)$ are monotonically increasing functions, since
$m' = \frac{1}{2} r^2 G_{00} \geq 0$ and
$S'/S = \frac{1}{2} r N^{-1} [G_{00} + G_{11}] \geq 0$.
Second, the local mass function $M(r)$, defined by
\be
M(r) \, \equiv \, - \frac{1}{8 \pi} \int_{S^2_r} \ast dk \, = \, 
mS \, + \, N S' r^2 \, - \, m' S r \, ,
\label{local-Ko}
\ee
assumes the values $M$ and $\frac{1}{4 \pi} \kappa {\cal A}$ for
$r \rightarrow \infty$ and $r = r_H$, respectively.
Using $m' \geq 0$, $S' \geq 0$, asymptotic flatness and
$N(r_H) = 0$, we now find
\be
M \, \geq \, lim_{r \rightarrow \infty} \left( m \, S \right) \, 
\; \; \; \; \mbox{and} \; \; \;  
\frac{1}{4 \pi} \kappa {\cal A} \, \leq \, m(r_H) \, S(r_H) \, .
\label{est}
\ee
Finally, taking again advantage of the monotonicity of $m$ and $S$, 
eq. (\ref{est}) yields the desired result.

As an application of the above uniqueness results
we consider scalar fields or, more generally,
{\em harmonic\/} mappings $\phi$ from spacetime $(M,\gtens)$ into 
{\em Riemannian\/}
target manifolds $(N,\Gtens)$. The action is
\be
{\cal A} \, = \, \int \left( - \frac{1}{16 \pi} R \, + \, 
\frac{1}{2} \sprod{d \phi}{d \phi}_G \, + \, P[\phi]
\right) \ast 1 \, ,
\label{H-A}
\ee
where we have also allowed for a nonnegative potential $P[\phi]$.
In local coordinates the harmonic density becomes
$\sprod{d \phi}{d \phi}_G \equiv G_{AB} 
\sprod{d \phi^A}{d \phi^B} =
G_{AB}[\phi(x)] g^{\mu \nu}(x) \phi^A,_{\mu} \phi^B,_{\nu}$.
(If the target manifold is a linear space, then this reduces to the
ordinary kinetic term for $\mbox{dim}(N)$ decoupled scalar fields.)
The important observation is that the stress-energy tensor,
\be
\Ttens \, = \, G_{AB}[\phi] \, d \phi^A \otimes d \phi^B \, - \, 
\gtens \, \left(
\frac{1}{2} \sprod{d \phi}{d \phi}_G \, + \, P[\phi] \right)
\label{se-tens}
\ee
violates the SEC in the above sense, since
\be
\Ttens(k,k) \, - \, \frac{1}{2} \, \gtens(k,k) \; \mbox{tr} \Ttens \, = \, 
- \, V \, P[\phi] \, \leq \, 0 \, ,
\label{se-tens2}
\ee
where we have required that $\phi$ is a stationary
mapping, $d \phi^A (k) = L_k \phi^A = 0$.
We are now able to conclude that selfgravitating harmonic mappings
with non-negative potentials and Riemannian target spaces admit\\
$\bullet$ no soliton solutions,\\
$\bullet$ no static, spherically symmetric black hole solutions other than
Schwarzschild.

In order to get rid of the requirement of spherical symmetry in the
black hole case, one needs either a proof for 
$M \geq \frac{1}{4\pi} \kappa {\cal A}$ or a different argument.
A very powerful alternative is provided by divergence identities, as
Bekenstein originally used in this context \cite{JB-DI}. 
This method gives, however, only conclusive results
for {\em convex\/} potentials and target spaces with 
{\em nonpositive\/} sectional curvature (see, e.g. \cite{GG}).
We shall return to this problem and its partial resolution
in sections 7 \& 9.

\section{Staticity and Circularity}

The integrability theorems for the Killing fields provide the link between 
the strong rigidity theorem and the assumptions on which the classical
uniqueness theorems are based.
In the nonrotating case one has to show that a stationary 
domain of outer communications (with Killing field $k$, say) 
is static, 
\be
\ast \left( k \we dk \right) \, = \, 0 \, ,
\label{F1}
\ee
whereas in the stationary and axisymmetric situation 
(with Killing fields $k$ and $m$, say) one must
establish the circularity conditions:
\be
\ast \left( m \we k \we dk \right) \, = \, 0 \, ,
\; \; \; \; 
\ast \left( k \we m \we dm \right) \, = \, 0 \, .
\label{F2}
\ee
(The Hodge duals are taken for later convenience.)
As mentioned in the introduction, the vacuum staticity 
and circularity theorems were proven by Carter and Hawking, 
respectively.
Whilest Carter also succeeded in establishing the electrovac 
circularity theorem in the early seventies (see, e.g. \cite{C-87}), 
it took, however, some effort until the corresponding staticity 
issue was settled as well \cite{SW-92-93}.
 
In this section we shall first show that -- under fairly
mild assumptions -- the Frobenius integrability conditions
(\ref{F1}) and (\ref{F2})
are equivalent to the (local) {\em Ricci conditions\/},
\be
\ast \left( k \we R(k) \right) \, = \, 0 \, ,
\label{R1}
\ee
and 
\be
\ast \left( m \we k \we R(k) \right) \, = \, 0 \, ,
\; \; \; \; 
\ast \left( k \we m \we R(m) \right) \, = \, 0 \, ,
\label{R2}
\ee
respectively. 
This equivalence will then enable us\\
$\bullet$
to prove the vacuum staticity theorem for not necessarily
connected horizons;\\
$\bullet$
to prove the staticity and circularity theorems for scalar
mappings (e.g. Higgs fields);\\
$\bullet$
to understand why the electrovac circularity theorem does {\em not}
generalize to YM fields.

We start by noting that the Ricci conditions
(\ref{R1}) and (\ref{R2}) are the derivatives of the 
Frobenius conditions (\ref{F1}) and (\ref{F2}), 
respectively: The staticity condition (\ref{F1}) is,
of course, equivalent to the vanishing of the twist
$\omega_k$ assigned to the stationary Killing field
($\omega_k \equiv \frac{1}{2}\ast [ k \we dk ]$). Hence,
\be
\frac{1}{2} \, d \ast \left( k \we dk \right)
\, = \, d \omega_k
\, = \, \frac{1}{2} \ast \left( k \we \cod dk \right)
\, = \, \ast \left( k \we R(k) \right) \, ,
\label{der1}
\ee
where $\cod \equiv \ast d \ast$, and where we have
used the Ricci identity for Killing fields in the last step
(see, e.g. \cite{MH-96-LN} for details). 
As for the derivatives of the circularity conditions (\ref{F2}), 
we note that $\ast (m \we k \we dk)$ can be written as 
$-2 i_m \omega_k \, (= -2 m_{\mu} \omega_k^{\mu}$). 
We therefore find
\be
\frac{1}{2} \, d  \ast \left( m \we k \we dk \right) \, = \,
- d \left( i_{m} \omega_{k} \right) \, = \, i_{m} d \omega_{k} \,
= \, i_{m} \ast \left( k \we R(k) \right) \, = \, \ast
[m \we k \we R(k)] \, .
\label{der2}
\ee
Here we have also used the fact that the Killing fields commute
\cite{C-70}. (See \cite{LS} for a powerful proof 
of $[k,m] = 0$ which does not require the existence of an axis.)

The above connection shows that the Frobenius conditions imply the
Ricci conditions. In the stationary and axisymmetric case, the
{\em converse\/} statement is also easily established, 
since the vanishing of the $1$-form $d(i_m \omega_k)$ implies
that the {\em function\/} $i_m \omega_k$ is constant. Since,
in addition, $i_m \omega_k$ vanishes on the rotation axis, the
circularity condition (\ref{F2}) follows from the Ricci condition 
(\ref{R2}) in every domain of spacetime intersecting the symmetry
axis. In the nonrotating case the situation is more difficult,
since $d \omega_k = 0$ does not automatically imply that the
twist $1$-form itself vanishes. However, using
$d (\frac{k}{V}) = \frac{2}{V^2} \ast (k \we \omega)$
(write $\ast(k \we \omega) = -i_k \ast \omega$ to obtain this),
it is not hard to prove the useful identity
\be
d \, \left( \omega \we \frac{k}{V} \right) \, = \, 
d \omega \we \frac{k}{V} \, - \, 
2 \frac{\sprod{\omega}{\omega}}{V^2} \ast k \, ,
\label{twist-id}
\ee
where, as usual, $V \equiv - \sprod{k}{k}$.
Since asymptotic flatness and the general properties of Killing horizons
imply that $\omega \we \frac{k}{V}$ vanishes at infinity and
at (each component of) the Killing horizon, we can apply Stokes' 
theorem for a spacelike hypersurface to conclude that the 
integral over the l.h.s. vanishes.
Hence, $d \omega = 0$ implies 
$\int_{\Sigma} V^{-2} \sprod{\omega}{\omega} \ast k = 0$ and 
therefore -- provided
that the domain of outer communications is {\em strictly\/} 
stationary  ($V \geq 0$) -- also $\omega = 0$. 
This shows that $\omega = 0$ follows from $d \omega = 0$
whenever the horizon is nonrotating and the domain is 
strictly stationary. In particular, no restrictions
concerning the connectedness of the horizon
enter the above argument. (The original proof of the
vacuum staticity theorem was based on the fact that $d \omega = 0$ 
implies the local existence of a potential, and was therefore 
subject to stronger topological restrictions.)
Hence, under the (weak) assumptions used above, it is
sufficient to establish the {\em Ricci\/} staticity and circularity
conditions (\ref{R1}), (\ref{R2}) 
in order to conclude that the {\em Frobenius\/} 
integrability conditions (\ref{F1}), (\ref{F2}) are satisfied. 

We shall now demonstrate that the circularity property is a 
consequence of the field equations for both 
scalar mappings and {\em Abelian\/} gauge fields, whereas the situation
is completely different for the EYM system.

\subsection{Scalar Mappings}

Consider a selfgravitating
mapping $\phi$ from spacetime $(M,\gtens)$ into a
target manifold $(N,\Gtens)$ with matter Lagrangian
${\cal L} = {\cal L}(\gtens, \Gtens, \phi, d \phi)$.
If ${\cal L}$ does not depend on derivatives of the
spacetime metric, then the stress-energy tensor becomes 
\be 
\Ttens \, = \, 2 \,  
\frac{\partial {\cal L}}{\partial \gtens ^{-1}} \, - \, 
{\cal L} \cdot \gtens \, = \, f_{AB} \, d \phi^A \otimes d \phi^B \, - \,
{\cal L} \cdot \gtens \, ,
\label{SES1}
\ee
where the $f_{AB} = f_{AB}(\gtens, \Gtens, \phi, d \phi)$ depend on
the explicit form of the Lagrangian. Let $\phi$ be invariant under
the $1$-parameter group of transformations generated by a Killing field
$\xi$, say, $L_{\xi} \phi = 0$. Then the stress-energy $1$-form with
respect to $\xi$, $T(\xi)$, becomes proportional to $\xi$ itself, 
$T(\xi) = - {\cal L} \cdot \xi$, and thus
\be
T(\xi) \, \we \xi \, = \, 0 \, .
\label{SES2}
\ee
Hence, a stationary mapping $\phi$ fulfills the Ricci staticity
condition (\ref{R1}), whereas the Ricci circularity conditions
(\ref{R2}) are satisfied for stationary and axisymmetric scalar mappings.
By virtue of the equivalence between these conditions and the
Frobenius integrability conditions (\ref{F1}) and (\ref{F2}), we
are able to conclude that stationary black hole spacetimes 
coupled to stationary scalar mappings are
either static or circular.

\subsection{Gauge Fields}

As mentioned earlier, the {\em staticity\/} theorem for Maxwell
fields has been established only recently \cite{SW-92-93}. 
In contrast to this, the corresponding circularity problem was 
solved by Carter already in the early seventies. 
Here we restrict ourselves to the {\em circularity\/} issue, 
which we will discuss for both electromagnetic {\em and\/} YM fields.
Our main objective is to understand why -- in contrast to the
Maxwell case -- circularity is not a consequence of the field equations 
when gravity is coupled to {\em non-Abelian\/} gauge fields. In order to 
do so, we first derive an expression for the
$2$--form $\xi \we \Txi$, where $\xi$ stands either for the stationary
or the axial Killing field.

We start by recalling that a gauge potential  
$A$ is symmetric with respect to the action of $\xi$ if there
exists a Lie algebra valued function $\psi$ (for every Killing field $\xi$),
such that 
\be
L_{\xi} A \, = \, D \psi
\label{K-Lie-A}
\ee
(see, e.g. \cite{FM}).
Using $L_{\xi} (dA + A \we A)$ $=$ $d L_{\xi} A + \commut{A}{L_{\xi} A}$
$=$ $D L_{\xi} A$ $=$ $D^{2} \psi$, the symmetry equation for the field
strength (and every other gauge-covariant $p$-form) becomes
\be
L_{\xi} F \, = \, \commut{F}{\psi} \, ,
\label{K-Lie-F}
\ee
(which, as expected, reduces to $L_{\xi} F = 0$ in the Abelian case).

It is convenient to introduce the electric, 
$E \equiv -i_{\xi} F$, and the magnetic, $B \equiv i_{\xi} \ast F$,
component of $F$ with respect to the Killing field $\xi$. (If $\xi$
denotes the timelike and hypersurface orthogonal Killing field 
of a {\em static\/} spacetime then $E$ and $B$ are the 
ordinary electric and magnetic $1$--forms.)
Since the inner product, $i_{\xi} \alpha$, of $\xi$ with 
an arbitrary $p$--form $\alpha$ is obtained from
$i_{\xi} \alpha = -\ast (\xi \we \ast \alpha)$, 
we have (with $\ast^{2} \alpha = -(-1)^{p} \alpha$)
\be
E \, = \, \ast \left( \xi \we \ast F \right) \, , \; \; \; \; \; 
B \, = \, \ast \left( \xi \we F \right) \, .
\label{K-EB}
\ee
An important consequence of the symmetry condition (\ref{K-Lie-A})
is the fact that -- for every Killing field $\xi$ -- there 
exists an {\em electric potential\/} $W$: Since 
$E = -i_{\xi} ( dA + A \we A )$ $=$ 
$-L_{\xi} A + d i_{\xi} A + \commut{A}{i_{\xi} A}$ $=$ 
$-D \psi + D (i_{\xi} A)$, one finds
\be
E \, = \, D W \, , \; \; \; \mbox{with} \; \; 
W \, \equiv \, i_{\xi}A \, - \, \psi \, .
\label{K-EW}
\ee
Now using the definitions (\ref{K-EB}) and some basic identities
for Killing fields (see, e.g. \cite{MH-96-LN})
it is not difficult to verify that the source-free YM equations
$D \ast F = 0$, together with the Bianchi identity $DF = 0$, are 
equivalent to the following set of equations
\be
\ast D \ast \left( \frac{E}{N} \right) \, = 
\, 2 \, \frac{\sprod{\omega}{B}}{N^2}
\, ,\; \; \; \; \; 
E \, = \, D W \, ,
\label{K-Eeqs}
\ee 
\be
\ast D \ast \left( \frac{B}{N} \right) \, = 
\, - \, 2 \, \frac{\sprod{\omega}{E}}{N^2} 
\, , \; \; \; \; \; 
DB \, = \, - \ast  DE \, ,
\label{K-Beqs}
\ee 
where $N$ and $\omega$ are the norm and the twist, respectively,
assigned to the Killing field $\xi$,
$N \equiv \sprod{\xi}{\xi}$,
$\omega = \frac{1}{2} \ast \left( \xi \we d \xi \right)$.

We are now able to express the $2$--form $\xi \we T(\xi)$ in terms of
the electric potential $W$ and the magnetic $1$-form $B$. In order to
do so, we use the YM stress-energy tensor
\be
T_{\mu \nu} \, = \, \frac{1}{4 \pi} \,
\trace{F_{\mu \sigma} F_{\nu}^{\; \; \sigma} \, - \, \frac{1}{2}
g_{\mu \nu} \sprod{F}{F}} \, ,
\label{FE-TYM}
\ee
and the above definitions for $E$ and $B$ to obtain
\be
\xi \we T(\xi) \, = \, \frac{1}{4 \pi} \ast \trace{E \we B} \, .
\label{K-2-YM}
\ee
Now using the YM equations $E = DW$, $DB = -\ast D^2 W = -\ast 
\commut{F}{W}$ and the fact that 
$\trace{W \commut{F}{W}}$ vanishes, we find
$\trace{E \we B}$ $=$ $\trace{DW \we B}$
$=$ $\trace{D(WB)} - \trace{W \/ DB}$ $=$
$d ( \trace{W \/ B} )$. We therefore obtain the result
\be
\ast \left( \xi \we \Txi \right) \, = \, 
- \frac{1}{4 \pi} \, d \, \left( \trace{W \, B} \right) \, = \, 
- \frac{1}{4 \pi} \, d \, \left( \trace{W \, i_{\xi} \ast F} \right) \, .
\label{ex-diff}
\ee
By virtue of the general identity (\ref{der1}) 
and Einstein's equations, the l.h.s. is 
(up to a factor of $\frac{1}{8\pi}$)
exactly the derivative of the twist assigned to $\xi$.
Since the r.h.s. is an exact differential-form as well, we
can integrate the equation and introduce
a ``combined twist potential'' $U$, such that
\be
dU \, = \, \omega \, + \, 2 \, \trace{W \/ B} \, .
\label{twpot}
\ee
Hence, we have expressed the twist in terms of an integration 
function $U$ and the gauge fields. 
For the stationary Killing field $k$,
the Frobenius staticity condition therefore becomes
\be
\omega_k \, = \, dU \, - \,  
2 \, \trace{W_k \, i_k \ast F} \, = \, 0 \, ,
\label{F-4}
\ee
where $W_k$ is the electric potential with respect to $k$; 
see eq. (\ref{K-EW}).

In a similar way one can integrate the right hand sides of the 
Ricci {\em circularity\/} conditions in order to obtain an expression
for the Frobenius conditions (\ref{F1}), (\ref{F2}). 
Since the latter can be written as
$i_m\omega_k = 0$ and $i_k \omega_m = 0$, we can also use
the above formula for the twist, which 
yields the desired expressions
\be
i_{m} \omega_k \, = \, -2 \, \trace{W_k \, (\ast F)(k,m)}
\, = \, 0 \, , \; \; \; \; 
i_{k} \omega_m \, = \, 2 \, \trace{W_m \, (\ast F)(k,m)}
\, = \, 0 \, ,
\label{circ-4}
\ee
where $(\ast F)(k,m) = i_m i_k \ast F$.
(Here we have used the circumstance that the constants of
integration vanish.)
Carter has derived the connection
between the circularity conditions and the gauge
fields in the Abelian case. His formula involves
magnetic potentials, which locally exist as a 
consequence of the {\em Maxwell} equations $d B_k = d B_m = 0$.
The above formulae (\ref{F-4}), (\ref{circ-4}) show that
even in the {\em non-Abelian} case -- where the magnetic
potentials do not exist any more -- one can 
integrate the Ricci conditions and obtain an explicit expression for
the Frobenius conditions in terms of the gauge fields.
(The same result can be established if Higgs fields
are taken into account as well.)

A very important peculiarity 
of the {\em Abelian\/} case is the  
fact  that both $F(k,m)$ and $(\ast F)(k,m)$ vanish as
a consequence of the symmetry properties and
Maxwell's equations. This is seen as follows: 
Since $k$ and $m$ commute, 
the operator $di_m i_k$ can be cast into the form
\be
d \, i_m i_k \, = \, i_k L_m \, - \, i_m L_k \, + \, 
i_m i_k \, d \, . 
\label{d-i-i}
\ee
Applying this to $F$ and $\ast F$, and
using the fact that the Hodge dual and the Lie derivative
with respect to a Killing field commute, we immediately find
(with $L_kF = L_mF = 0$)
\be
d \left[ F(k,m) \right] \, = \, i_m i_k \/ dF \, = \, 0 \, ,
\; \; \; \; 
d \left[(\ast F)(k,m) \right] \, = \ i_m i_k \/ d \ast F   \, = \, 0 \, .
\label{abel}
\ee
Hence, the functions $F(k,m)$ and $(\ast F)(k,m)$
are constant and vanish on the
rotation axis. We therefore have $F(k,m)= 0$ and 
$(\ast F)(k,m) = 0$ where,
by virtue of eq. (\ref{circ-4}), the latter
property implies
that the Frobenius circularity conditions are fulfilled.
This is Carter's electrovac circularity theorem, establishing
the integrability property for the Killing fields on the
basis of the Maxwell equations and the symmetry properties
of the gauge fields.

In contrast to this, circularity does {\em not\/} follow from the 
field equations in the non-Abelian case. That is, the quantities
$\trace{W_k \/ (\ast F)(k,m)}$ and
$\trace{W_m \/ (\ast F)(k,m)}$
do not vanish as a consequence of $DF = 0$, 
$D \ast F = 0$ and the symmetry equations $L_{k} A = D \psi_{k}$,
$L_{m} A = D \psi_{m}$. In particular, the diagonal boxes of
the fieldstrength tensor are no longer automatically zero. 
Instead of eq. (\ref{abel}) we now have
\bea
D \left[ F(k,m) \right] & = &
\commut{E_{k}}{W_{m}} \, - \, \commut{E_{m}}{W_{k}} \, ,
\nonumber \\
D \left[ (\ast F)(k,m) \right] & = &
\commut{B_{m}}{W_{k}} \, - \, \commut{B_{k}}{W_{m}} \, .
\label{non-abel}
\eea
(Use eq. (\ref{d-i-i}) 
and $L_{m} W_{k} = \commut{W_{k}}{\psi_{m}}$ to derive this. 
Also note that $F(k,m) = \commut{W_{k}}{W_{m}}$, since
$i_{m} i_{k} F = -i_{m} E_{k} = -i_{m} DW_{k}$ $=$
$-L_{m} W_{k} - \commut{i_{m}A}{W_{k}}$ $=$
$- \commut{W_{k}}{\psi_{m}} + \commut{W_{k}}{i_{m}A}$ $=$
$\commut{W_{k}}{W_{m}}$.)

The above arguments also imply that 
the usual $(2 \! + \! 2)$-split of the metric 
of a stationary and axisymmetric spacetime imposes 
additional constraints on the YM fields. 
Writing the metric in the circular Papapetrou
form and considering the corresponding ansatz for the 
gauge potential,
\bdm
\gtens \, = \, g_{tt} \, dt \otimes dt 
       \, + \, 2 \, g_{t \varphi} \, dt \otimes d \varphi
       \, + \, g_{\varphi \varphi} \, d \varphi \otimes d \varphi
 \, + \, \tilde{\gtens} \, ,
\edm
\be
A \, = \, W_{k} \, dt \, + \, 
W_{m} \, d \varphi \, + \, \tilde{A} \, ,
\label{gA}
\ee
the integrability conditions require that the scalar
$\tilde{\ast} \tilde{F}$ is orthogonal to $W_{k}$ and $W_{m}$
with respect to the inner algebra product,
\be
\trace{W_{\xi} \, \tilde{\ast} \tilde{F}} \, = \, 0 \, ,
\; \; \; \mbox{for} \; \xi \, = \, k, \/m \, . 
\label{og}
\ee
A detailed discussion of the circularity issue for
non-Abelian gauge fields will be
presented elsewhere \cite{BH-97}.

\section{The No-Hair Theorem for Harmonic Mappings}

Selfgravitating scalar mappings from spacetime $(M,\gfour)$
into Riemannian manifolds $(N,\Gtens)$ admit both
soliton and black hole solutions with hair. A particular
example is provided by the Skyrme model \cite{DHS-91-PLB}, 
which can be described as a static mapping
into $S^{3}$ \cite{MA-RU} with matter action
$\int \left( e_{2}[\phi] + e_{4}[\phi] \right) \ast 1$,
where $e_{2}[\phi] \equiv \frac{1}{2} 
G_{AB}[\phi]\sprod{d \phi^{A}}{ d \phi^{B}}$
is the harmonic part of the action and
$e_{4}[\phi] \equiv \frac{1}{4} G_{AC} G_{BD} 
\sprod{d \phi^{A} \we d \phi^{B}}{d \phi^{C} \we d \phi^{D}}$.
If, however, only the {\em harmonic\/} part 
is taken into account, then the situation
changes dramatically and one obtains the following 
uniqueness theorem:
\newpage
\noindent
The Kerr metric is the unique solution with nondegenerate 
Killing horizon ($\kappa \neq 0$) amongst all stationary
black hole solutions of selfgravitating {\em harmonic\/}
mappings into {\em Riemannian\/} target manifolds $(N,\Gtens)$,
with action
\be
{\cal A} \, = \, \int \left( - \frac{1}{16 \pi} R \, + \, 
\frac{1}{2} \/ G_{AB}[\phi] \/ 
\sprod{d \phi^{A}}{d \phi^{B}} \right) \ast 1 \, .
\label{H-A-2}
\ee

The proof of this theorem involves the following
three main steps: First, one has to show that all stationary
black hole solutions to the above action are either static
($k \we dk = 0$)
or circular
($m \we k \we dk = k \we m \we dm = 0$). 
In section 6 we have argued that this
is indeed the case, i.e., that the Killing fields fulfill
the integrability conditions.
It therefore remains to prove that\\
$\bullet$ 
there are no static black hole solutions other than 
Schwarzschild,\\
$\bullet$
there are no circular black hole solutions other than Kerr.\\
Under the additional assumption of spherical symmetry 
we have already presented a proof of the first assertion 
in section 5 -- playing off the strong and the
dominant energy conditions against each other.
Here we shall only give a brief outline of the rationale leading to
the above results, and refer to \cite{MH-96-LN} for details.

\subsection{The Static Case}

With respect to the static metric
\be
\gfour \, = \,  -S^2 \/ dt^2 \, + \, \gtens \, ,
\label{stat-met}
\ee
the Einstein equations obtained from the action
(\ref{H-A-2}) become
\be
\Delta^{(g)} S \, = \, 0 \, ,
\; \; \; \; 
R^{(g)} \, = \, \kappa \, G_{AB} \, g^{ij} \, 
\phi^A ,_{i} \phi^B ,_{j} \, ,
\label{f-1}
\ee
\be
R^{(g)}_{ij} \, - \, 
S^{-1} \nabla_{j}^{(g)} \nabla_{i}^{(g)} S \, = \, \kappa \, G_{AB} \, 
\phi^A ,_{i} \phi^B ,_{j} \, ,
\label{f-2}
\ee
where Latin indices refer to the spatial metric $\gtens$.
In addition, we have the matter equations for the scalar fields
$\phi^A$, 
$\Delta^{(g)} \phi^A$ $+$ 
$S^{-1} (dS | d \phi^A)$ $+$
$\Gamma^A_{BC}(\phi) (d \phi^B | d \phi^C)$ $=$ $0$.
Like in the new proof of the static vacuum \cite{BM-87} 
and electrovac \cite{Ru} \cite{MuA-93}
uniqueness theorems, the strategy is to show that the
$3$-dimensional spacelike manifold
$\Sgfold$ is conformally flat.
In fact, the conformal factor 
which yields the desired result is the same as in 
the vacuum case, 
\be
\Omega_{\pm} \, = \, \frac{1}{4} (1 \pm S)^2 \, .
\label{su-1}
\ee
More precisely, we consider the manifold 
$\hat{\Sigma}$ which is composed of the
two copies $\hat{\Sigma}_{+}$ and $\hat{\Sigma}_{-}$ of $\Sigma$, 
pasted together along their boundaries
${\cal H}_+$ and ${\cal H}_-$. Furnishing
$\hat{\Sigma}_{+}$ and $\hat{\Sigma}_{-}$ with the metrics
\be
\ghattens_{\pm} \, = \, \Omega^2_{\pm} \, \gtens \, ,
\label{su-2}
\ee
one can show that the metric and the second fundamental form 
match continuously on ${\cal H}_{\pm}$. Moreover,
$(\hat{\Sigma}_+,\hat{\gtens}_+)$ 
is asymptotically flat with vanishing mass and
non-negative Ricci curvature, $\hat{R} \geq 0$:
Using the expansions 
$\gtens = (1 + \frac{2M}{r}) \deltens + {\cal O}(r^{-2})$ and
$\Omega_{+} = 1 - \frac{M}{r} + {\cal O}(r^{-2})$ one finds
$\ghattens_{+} = \Omega_{+} \gtens = 1 + {\cal O}(r^{-2})$,
which shows that 
$(\hat{\Sigma}_+,\hat{\gtens}_+)$ 
has vanishing mass. Computing the
conformal transformation of the Ricci scalar gives 
\bea
\frac{\Omega^{\, \, 4}_{\pm}}{2}\, \hat{R} & = &
\frac{\Omega^{\, \, 2}_{\pm}}{2} \, R^{(g)} \, - \, 
2 \, \Omega_{\pm} \, \Delta^{(g)} \Omega_{\pm} \, + \,
\sprod{d \Omega_{\pm}}{d \Omega_{\pm}}^{(g)}
\nonumber \\ & = &
\frac{\Omega^{\, \, 2}_{\pm}}{2} \, R^{(g)} \, - \,
\Omega_{\pm} \, (S \pm 1) \Delta^{(g)} S \, = \,
\frac{\Omega^{\, \, 2}_{\pm}}{2} \, \kappa
\sprod{d \phi}{d \phi}_{G} \, \geq \, 0 \, ,
\label{su-3}
\eea
where we have used eqs. (\ref{f-1}) and the
fact that the target metric $\Gtens$ is Riemannian.

Since ($\hat{\Sigma},\hat{\gtens}$) 
constructed in this way is an asymptotically flat,
complete, orientable $3$-dimensional Riemannian manifold with
non-negative Ricci curvature and vanishing mass, it is -- as a
consequence of the positive mass theorem -- isometric to
$(I \! \! R^{3}, \deltens)$. Hence, the spatial geometry
of the domain of outer communications is conformally flat,
$\hat{R} = 0$, which implies that the scalar fields assume constant
values. The field equations (\ref{f-1}), (\ref{f-2}) 
therefore reduce to the vacuum Einstein equations.
Finally, the fact that the static vacuum solutions
with conformally flat $3$-geometry are spherically symmetric 
concludes the proof of the static no-hair theorem for harmonic
mappings \cite{MH-93}.

\subsection{The Stationary and Axisymmetric Case}

Like in the static situation, the uniqueness problem for
stationary and axisymmetric black hole solutions
of selfgravitating harmonic mappings can be reduced
to the corresponding vacuum problem.
This is due to the following observations:\\
$\bullet$
First, the twist is closed, since {\em harmonic\/} mappings
have the special property that the Ricci
$1$-form (with respect to either of the Killing fields) vanishes,
\be
R(k) \, = \, R(m) \, = \, 0 \, .
\label{spec-harm}
\ee
(Use $L_k \phi^A = L_m \phi^A = 0$ and eq. (\ref{se-tens}) 
with $P[\phi] = 0$ to obtain this.)
One can therefore introduce the same Ernst potential as
in the vacuum case,
\be
\Erpot \, = \, - \, X \, + \, i \/ Y \, ,
\; \; \; \; \mbox{where} \; \; 
X \, \equiv \, \sprod{m}{m} \, , \; \; 
dY \, \equiv \, 2 \omega \, = \, \ast(m \we dm) \, .
\label{Ernst}  
\ee
$\bullet$
Second, the general identity
$X \Delta X - \sprod{dX}{dX} + 4 \sprod{\omega}{\omega} + 2XR(m,m) = 0$
reduces to the following equation for $\Erpot$
\be
\frac{1}{\rho} \, d^{\dagger (\gamma)} (\rho \, d\Erpot) \, = \, 
\frac{1}{X} \, \sprod{d\Erpot}{d\Erpot}^{(\gamma)} \, .
\label{EE}
\ee
This is again a consequence of $R(m) = 0$ and the fact that a
circular spacetime admits a foliation by
$2$-dimensional integrable surfaces orthogonal to the Killing fields
$k$ and $m$, implying that the metric can be written in the 
Papapetrou $(2\!+\!2)$ form
\be
\gfour \, = \, \sigtens \, + \, \frac{1}{X} \gatens \, = \,
- \, \frac{\rho^2}{X} \, dt^2
\; + \; X \, (d\varphi \, + \, A \, dt )^2 \; + \; \frac{1}{X} \,
\gatens \, ,
\label{metric-circ}
\ee
where the function $A$ is related to the twist potential $Y$ by
$dA = \rho X^{-2} \ast^{(\gamma)} dY$.
\newpage
\noindent
$\bullet$ Third, the determinant 
$\rho = \sqrt{- \mbox{det}(\sigtens})$ 
is subject to the equation
$X \rho \Delta^{(\gamma)} \rho$ $=$ 
$\sprod{k}{k} R(m,m)$ $+$ $\sprod{m}{m} R(k,k) - 2 \sprod{k}{m} R(k,m)$
which, like in the vacuum case, reduces to 
\be
\Delta^{(\gamma)} \rho \; = \; 0 \, .
\label{harm-rho} 
\ee
This shows that $\rho$ is a harmonic function on the $2$-dimensional
Riemannian manifold $(\Gamma, \gatens)$. One can therefore
introduce Weyl coordinates, that is, one can use
$\rho$ and its conjugate harmonic function $z$, say,
as coordinates on $(\Gamma, \gatens)$
(see \cite{GW} for an elegant and complete proof). 

The above observations imply that eq. (\ref{Ernst}) for the 
twist and the norm of the axial Killing field 
decouples from the remaining field equations and is, in fact,
identical to the {\em vacuum\/} Ernst equation,
\be
\frac{1}{\rho} \, \nabdel(\rho \, \nabdel \Erpot ) \, + 
\, \frac{(\nabdel \Erpot | \nabdel \Erpot)}{X} \, = \, 0 \, ,
\label{Vac-Ernst}
\ee
where $\nabdel = (\partial_{\rho},\partial_{z})$.
Since the latter describes a regular boundary-value problem
on a fixed $2$-dimensional background space, 
one can apply the vacuum
Mazur \cite{Maz-82-84b} (Robinson \cite{DCR-75}) identity
to conclude that the Ernst potential is uniquely
determined by the boundary conditions. Also using 
$\frac{1}{\rho} A,_{\rho} = X^{-2} Y,_z$ and
$\frac{1}{\rho} A,_z = -X^{-2} Y,_{\rho}$ shows that the
metric $\sigtens$ of the orbit manifold is identical with the
corresponding vacuum solution.
It remains to discuss the orthogonal set of equations for the
scalar fields and the metric $\gatens$.

Like the Ernst equation, the field equations for the harmonic
scalar field $\phi$,
\be
\frac{1}{\rho} \, \nabdel(\rho \, \nabdel \, \phi^A) \, + 
\, \Gamma^A_{\; BC} \; (\nabdel \phi^B | \nabdel \phi^C) \, = \, 0 \, ,
\label{Matter-Ernst}
\ee
involve no unknown metric functions. 
The only remaining metric function $h(\rho,z)$ -- defined by
$\gatens = e^{2h} (d \rho^2 + dz^2)$ -- is therefore obtained 
by quadrature from the Ernst potential $\Erpot$ and the solution 
$\phi$ to the above matter equation: Writing
$h = h^0 + 8 \pi G \/ \tilde{h}$, it remains to integrate 
\be
\frac{1}{\rho} h^{0},_{\rho} = \frac{1}{4 X^2}
\left[ \Erpot,_{\rho} \bar{\Erpot},_{\rho} - 
\Erpot,_z \bar{\Erpot},_z \right] \, , \; \; \; 
\frac{1}{\rho} h^{0},_z = \frac{1}{4 X^2}
\left[\Erpot,_{\rho} \bar{\Erpot},_z  + 
\Erpot,_z \bar{\Erpot},_{\rho} \right] 
\label{h-0}
\ee
and
\be
\frac{1}{\rho} \tilde{h},_{\rho} = \frac{G_{AB}[\phi]}{2}
\left[ \phi^A,_{\rho} \phi^B,_{\rho} -
\phi^A,_z \phi^B,_z \right] \, , \; \; \; 
\frac{1}{\rho} \tilde{h},_z = \frac{G_{AB}}{2}
\left[ \phi^A,_{\rho} \phi^B,_z +
\phi^A,_z \phi^B,_{\rho} \right] \, .
\label{h-1}
\ee
The last step in the uniqueness proof is to show that the matter 
equations (\ref{Matter-Ernst}) and (\ref{h-1}) 
admit only the trivial solution $\phi^A = \phi^A_0$, $\tilde{h} = 0$.
Using Stokes' theorem, this is established as a
consequence of asymptotic flatness, the
fall-off conditions
\be
\phi^A = \phi^A_{\infty} + {\cal O}(r^{-1}) \, , \; \; \; 
\phi^A,_{r} = {\cal O}(r^{-2}) \, , \; \; \;
\phi^A,_{\vartheta} = {\cal O}(r^{-1})
\label{fall-off}
\ee
and the requirement that $G_{AB}[\phi]$ and the derivatives of the 
scalar fields with respect to the 
Boyer-Lind\-quist coordinates $r$ and $\vartheta$ remain finite at 
the boundary of the domain of outer communications
(see \cite{H-ROT} for details). This finally 
establishes the uniqueness of the Kerr metric amongst the 
stationary and axisymmetric black hole solutions of selfgravitating
harmonic mappings.

\section{Integral Identities}

Integral identities play an outstanding role in the derivation
of various uniqueness results. In fact, both Israel's 
uniqueness theorem for nonrotating vacuum black holes \cite{WI-67}
and Robinson's theorem for the corresponding rotating situation
\cite{DCR-75} were based on ingenious applications of
Stokes' law. Moreover, the fact that the electric potential
depends only on the gravitational potential -- a result which
opened the way for the static electrovac uniqueness 
theorem -- was also obtained by this technique \cite{WI-68}.
Finally, it was a divergence identity for sigma-models
on symmetric spaces which provided the key to
the uniqueness proof for the
Kerr-Newman metric \cite{Maz-82-84b}, \cite{Bunt-83}.
Moreover, many nonexistence results for {\em soliton}
configurations are also based on integral formulae.
For instance, the integrated version of the Ricci identity
for a stationary Killing field, $d \ast dk = 2 \ast R(k)$,
provides the link between the total mass, 
$M = -8 \pi G \int_{\infty} \ast dk$,
and the volume integral over the stress-energy $1$-form
$\Ttens(k) - \frac{1}{2} \mbox{tr} (\Ttens) \/ k$.
In combination with the positive mass theorem, this excludes
the existence of purely gravitational solitons other than
flat spacetime.

For spacetimes admitting a stationary -- not necessarily 
static -- isometry (with Killing field $k$, say), 
it turns out to be convenient to use the following version 
of Stokes' theorem: Consider a $1$-form $\alpha$ which is
invariant under the action of $k$, $L_k \alpha = 0$.
Then, using the fact that the Lie derivative with respect
to a Killing field commutes with the Hodge dual, one finds
$0 = \ast L_k \alpha = L_k \ast \alpha$ $=$ 
$i_k d \ast \alpha + d i_k \ast \alpha$. Since $\cod \alpha$
is a $0$-form, we have
$i_k d \ast \alpha = - (\cod \alpha) i_k \eta$ and
$i_k \ast \alpha = -\ast (k \we \alpha)$. 
Hence, integrating over a spacelike hypersurface $\Sigma$ with
boundary $\partial \Sigma$, Stokes' formula becomes
\be
\int_{\partial \Sigma} \ast \/ \left( k \we \alpha \right) \, = \, 
- \, \int_{\Sigma} ( \cod \alpha ) \, i_k \eta \, .
\label{Stokes-1}
\ee
Using the field equations (matter or gravitational)
for $\cod \alpha$ one now obtains, for instance, 
relations between asymptotic charges and quantities defined
on the horizon. In the best case one can combine the
field equations such that the sum of the boundary terms on the l.h.s.
vanishes and the sum of the integrands on r.h.s. has a fixed sign.
The method will be explained in the last section for
the Einstein-Maxwell equations. A well-known
example of this kind is Bekenstein's no-hair theorem for
scalar fields.

\section{The Bekenstein Argument}

In 1972 Bekenstein showed that there are no black holes with nontrivial
massive scalar fields \cite{JB-DI}. His strategy was to multiply the
matter equation $( \Delta - m^2) \phi = 0$ by $\phi$, integrating by
parts and using Stokes' theorem for the resulting identity. More
generally, one can consider an arbitrary function $f[\phi]$ of the scalar
field and choose $\alpha = f \/ d \phi$ in eq. (\ref{Stokes-1}).
The scalar field equation (with potential $P[\phi]$),
\be
\Delta \phi \, = \, - \, \cod d \phi \, = \, P,_{\phi} \, ,
\label{Bek-1} 
\ee
then yields
\bdm
\cod \alpha \, = \, \cod \left( f \, d \phi \right) \, = \, 
- \, \sprod{df}{d \phi} \, + \, f \, \cod d \phi \, = \, - \, f,_{\phi}
\sprod{d \phi}{d \phi} \, - \, f \/ P,_{\phi} \, .
\edm
Now using Stokes' theorem (\ref{Stokes-1}) and the fact that 
the scalar field assumes a constant value on 
the horizon and at spacelike infinity, we find
\be
4 \pi \, \left( f[\phi_{\infty}] \, s_{\infty} \, - \, 
f[\phi_H] \, s_H \right) \, = \,
\int_{\Sigma} \left( 
f,_{\phi} \sprod{d \phi}{d \phi} \, + \, f \, P,_{\phi} \right)
\, i_k \eta \, ,
\label{Bek-2}
\ee
where we have defined $s_{\infty \, (H)}
\equiv \frac{1}{4 \pi} \int_{S^2_{\infty} \, (H)} \ast (k \we d \phi)$.
Using general properties of Killing horizons, it is not difficult to see
that $s_H$ vanishes. Hence, provided that $f[\phi_H]$ is finite, the
horizon term on the l.h.s. does not contribute.  Our aim is to
choose $f[\phi]$ such that the asymptotic term vanishes as well. Since
asymptotic flatness implies that the scalar charge $s_{\infty}$ is
finite, it is sufficient to require 
$f[\phi_{\infty}] = 0$. Following Bekenstein
\cite{JB-DI}, we set $f[\phi] = \phi - \phi_{\infty}$. The above
formula then gives
\be
\int_{\Sigma} \left[ 
\sprod{d \phi}{d \phi} \, + \, (\phi - \phi_{\infty}) 
\, P,_{\phi} \right]
\, i_k \eta \, = \, 0 \, , \; \; \; \; 
\mbox{if $|\phi_H - \phi_{\infty}| < \infty$},
\label{Bek-3}
\ee
where we recall that the horizon contribution was argued away only for
$|f[\phi_H]| < \infty$. Considering a convex, non-negative potential with
$P,_{\phi}(\phi_{\infty}) = 0$, 
the integrand is non-negative and the above relation
shows that $\phi$ must assume the constant value 
$\phi_{\infty}$. (Also note that $d \phi$ is nowhere
timelike since $\sprod{k}{d \phi} = L_k \phi = 0$ 
and since the domain is
required to be strictly stationary.)
It is worth pointing out that the
Bekenstein black hole solution for a free scalar 
field \cite{JB-BH} does not contradict the above argument, 
since the latter applies only to scalar
fields which remain finite on the horizon. 

A slightly different version of the above result is obtained by
setting $f[\phi] = P,_{\phi}$, i.e., $\alpha = f \/ d \phi = dP$
\cite{MH-96-LN}.
Using again asymptotic flatness and the fact that the curvature invariant
$R_{\mu \nu} R^{\mu \nu}$ must remain finite on the horizon
enables one to conclude that
\be
\int_{\Sigma} \left[ P,_{\phi \phi} \sprod{d \phi}{d \phi}
\, + \, (P,_{\phi})^2 \right] \, i_k \eta \, = \, 0 \, .
\label{Bek-4}
\ee
This shows that there are only trivial scalar field configurations
for black hole solutions with nondegenerate, nonrotating
horizon and strictly stationary domain of outer communications.

I am not aware of a uniqueness theorem for the Schwarzschild metric
for scalar fields with {\em arbitrary\/}, non-negative 
potentials -- unless spherical symmetry is
imposed (see sections 4 \& 5).
In this connection, it is probably worth
noticing that the Bekenstein argument makes no essential use
of Einstein's equations. In an attempt to include more information
one can, for instance, take the $R_{00}$ equation into account as well,
\be
\cod \left( \frac{dV}{V} \right) \, = \, - \frac{2}{V} \, R(k,k) \, = \,
2 \kappa \, P[\phi] \, .
\label{BB-6}
\ee
Choosing $\alpha = g(\phi , V) \, \frac{dV}{V}$ and 
$\alpha = f(\phi , V) \, d \phi$ in Stokes' theorem
(\ref{Stokes-1}) yields the two integral equations
\bea
8 \pi \, \left( g_{\infty} \, M \, - \, 
g_H \, \frac{\kappa {\cal A}}{4 \pi} \right) & = & 
\int_{\Sigma} 
\left[ g,_V \, \frac{\sprod{dV}{dV}}{V} \, + \,
g,_{\phi} \, \frac{\sprod{d \phi}{dV}}{V} \, - \,
\kappa \, g \, P \right] \, i_k \eta \, ,
\nonumber \\
4 \pi \, \left( f_{\infty} \, s_{\infty} \, - 
\, f_H \, s_H \right) & = &
\int_{\Sigma} 
\left[ f,_{\phi} \sprod{d \phi}{d \phi} \, + \, 
f,_{V} \sprod{dV}{d \phi} \, + \, 
f \, P,_{\phi} \right]
\, i_k \eta \, ,
\label{Bek-7}
\eea
where we have also used the Komar formula, the identity 
$\ast(k \we \frac{dV}{V}) = - \ast dk -2 \frac{k}{V} \we \omega$ and
the fact that the boundary integral of $\frac{k}{V} \we \omega$
vanishes. 
One can now try to combine these formulae in order to
derive relations between the boundary terms. A proof of the
scalar uniqueness theorem for arbitrary non-negative
potentials along these or other lines would close a longstanding gap. 

The technique presented in this section can also be applied to
selfgravitating harmonic mappings (generalized scalar fields).
This yields uniqueness theorems for mappings into 
target manifolds with nonpositive sectional curvature
(see \cite{GG}). 
Since this restriction was absent in the uniqueness proof
presented in section 7.1., the conformal techniques are, in this case,
more powerful than the divergence identities.

\section{The Israel Theorem}

This celebrated theorem establishes that all static black hole solutions
of Einstein's vacuum equations are spherically symmetric \cite{WI-67}.
Israel was able to obtain this result -- and its extension
to electrovac spacetimes -- by considering a particular
foliation of the static $3$-dimensional hypersurface $\Sigma$:
Requiring that $S \equiv \sqrt{-\sprod{k}{k}}$ is an admissible
coordinate (see also \cite{MZH-73-74}), the spacetime metric can be
written in the $(1\!+\!1\!+\!2)$-split
\be
\gfour \, = \, -S^2 \, dt \otimes dt \, + \, 
\rho^2 \, dS \otimes dS \, + \, \gtiltens \, ,
\label{Is-split}
\ee
where both $\rho$ and the metric $\gtiltens$ depend on $S$ and the
coordinates of the $2$-dimensional surfaces with constant $S$.  With
respect to the tetrad fields $\theta^0 = S dt$ and $\theta^1 = \rho dS$
one finds
\be
G_{00} + G_{11} \, = \,
\frac{1}{\rho}  \left[ \frac{K}{S} - \frac{\partial K}{\partial S} - 
\frac{\rho}{2} K^2 \right]
\, - \, \frac{2}{\sqrt{\rho}} \, \laptil \sqrt{\rho}
\, - \, 
\left[ \frac{\sprod{\nabtil \rho}{\nabtil \rho}}{2 \rho^2}
\, + \, \tfree{K} _{ab} \tfree{K} ^{ab} \right] \, ,
\label{Is1}
\ee
\be
G_{00} + 3 \, G_{11} \, = \,
\frac{1}{\rho} \left[ 3 \/ \frac{K}{S} \, - \, 
\frac{\partial K}{\partial S} \right] - \tilde{R} \, - \,   
\laptil \ln \rho \, - \, 
\left[ \frac{\sprod{\nabtil \rho}{\nabtil \rho}}{\rho^2}
\, + \, 2 \tfree{K} _{ab} \tfree{K} ^{ab} \right] \, ,
\label{Is2}
\ee
where $K_{ab} = -(2 \rho)^{-1} \partial \tilde{g}_{ab}/ \partial S$ is
the extrinsic curvature of the embedded surface $S= const.$ in $\Sigma$,
and $\tfree{K} _{ab} \equiv K_{ab} - \frac{1}{2} \tilde{g}_{ab} K$ is the
tracefree part of $K_{ab}$.  Using the Poissson equation
\be 
\frac{\partial \rho}{\partial S} \, = \, \rho^2 \, \left( 
K \, - \, \rho \/ S \, R_{00} \right) \, ,
\label{Is3}
\ee
and the vacuum Einstein equations, one obtains
the following inequalities from
eqs. (\ref{Is1}) and (\ref{Is2}):
\be
\frac{\partial}{\partial S} \, 
\left( \frac{\sqrt{\gtil}}{\sqrt{\rho}}
\, \frac{K}{S} \right) \, \leq \,
-2 \, \frac{\sqrt{\gtil}}{S} \, \laptil \sqrt{\rho} \, ,
\; \; \; \; \;
\frac{\partial}{\partial S} \, 
\left( \frac{\sqrt{\gtil}}{\rho} \, [ KS + \frac{4}{\rho} ] \right)
\, \leq \,
- \, S \, \sqrt{\gtil} \, ( \laptil \ln \rho + \tilde{R} ) \, ,
\label{Is6}
\ee
where equality holds if and only if $\tfree{K} _{ab} = \nabtil \rho =
0$.  The strategy is now to integrate the above estimates over a
spacelike hypersurface extending from the horizon ($S=0$) to
$S^2_{\infty}$ ($S=1$).  Using the Gauss-Bonnet theorem, $\int_S
\tilde{R} \, \tilde{\eta} = 8 \pi$, one finds
\be 
\left[ \, \int_S \frac{K}{\sqrt{\rho} S} \, \tilde{\eta}
\, \right] _{\, 0}^{\, 1} \, \leq 0 \, , \; \; \; \; 
\left[ \, \int_S \frac{K S + 4 \rho^{-1}}{\rho} \, \tilde{\eta}
\, \right] _{\, 0}^{\, 1} \, \leq - 4 \pi \, .
\label{Is13} 
\ee 
Finally, taking advantage of asymptotic flatness and the fact that on
the horizon $K_{ab} = 0$ and $K/S = \rho \tilde{R} / 2$, one obtains $8
\pi \sqrt{M} - 4 \pi \sqrt{\rho_H}$ for the first integrand, and $- 4
{\cal A} \rho^{-2}_H$ for the second one (where ${\cal A}$ denotes the
area of the horizon).
Since $1/\rho_H$ equals the surface gravity of the horizon,
$1/\rho_H = \kappa$, the above inequalities eventually
give the estimates
\be
M \, \leq \, \frac{1}{4 \kappa} \, ,
\; \; \; \; \;
\frac{1}{4 \pi} {\cal A} \, \kappa \,
\geq \, \frac{1}{4 \kappa} \, ,
\label{Is14}
\ee
where we recall that equality holds if and only if both
$\tfree{K} _{ab}$ and $\nabtil \rho$ vanish.
Since the Komar expression for the total mass of a static {\em vacuum\/}
black hole spacetime yields $M = \frac{1}{4 \pi} \kappa {\cal A}$
(see eq. (\ref{Smarr})), we conclude that equality must hold
in the above estimates. This is, however, equivalent to
\be 
K_{ab} - \frac{1}{2} \gtil_{ab} K \, = \, 0 \, ,
\; \; \; \; \; 
\nabtil \rho \, = \, 0 \, .
\label{Is16}
\ee
Using this in eqs. (\ref{Is1}) and (\ref{Is3}) shows that both
$\rho$ and $K$ depend only on $S$. Equation (\ref{Is2}) then
implies that $\tilde{R}$ is constant on the surfaces of constant $S$.
Explicitly one finds
$\rho = 4 c \/ (1-S^2)^{-2}$, 
$K = c^{-1} S (1-S^2)$ and
$\tilde{R} = \frac{1}{2} c^{-2} (1-S^2)^2$,
where $c$ is a constant of integration.
Hence, defining
$r(S)$ by the relation $\tilde{R} = 2/r^2$, yields (with $c = M$)
\be
S^2 \, = \, 1-\frac{2 M}{r} \, ,
\; \; \; \;
\rho^2 dS^2 \, = \, (1-\frac{2 M}{r})^{-1} dr^2 \, ,
\; \; \; \;
\tilde{\gtens} = r^2 d \Omega^2 \, ,
\label{Is-2bis}
\ee
which is the familiar form of the Schwarzschild metric.

\section{The Mazur Identity}

Before we give an outline of Mazur's uniqueness proof for
the Kerr-Newman metric, we briefly recall some basic facts
about the Einstein-Maxwell equations with a Killing field.
We consider the axial Killing field $m$ with norm
$X = \sprod{m}{m}$ and twist 
$\omega_{m} = \frac{1}{2} \ast (m \we dm)$.
Together with the symmetry condition $L_{m} F = 0$ for the
electromagnetic field tensor ($2$-form) $F$, the Bianchi 
identity ($dF = 0$) and the Maxwell
equation ($d \ast F = 0$) imply that there exist two potentials
$\phi$ and $\psi$, such that $d \phi = -i_{m} F$ and
$d \psi = i_{m} \ast F$. In addition, the identity
$d \omega_{m} = \ast [m \we R(m)]$
implies $d \omega_{m} = -2 d \phi \we d \psi$ and thus 
the existence of a twist potential $Y$, such that
$\frac{1}{2} dY = \omega_{m} + \phi d \psi - \psi d \phi$.
The $R(m,m)$ Einstein equation,
$\cod(\frac{dX}{X})$ $-$ 
$\frac{4}{X^{2}} \sprod{\omega_{m}}{\omega_{m}}$ $=$
$\frac{2}{X} (\sprod{d \phi}{d \phi} + \sprod{d \psi}{d \psi})$,
and the remaining Maxwell equations,
$\cod(\frac{d\phi}{X})$ $=$ $\frac{2}{X^{2}} \sprod{\omega_{m}}{d \psi}$
and
$\cod(\frac{d\psi}{X})$ $=$ $-\frac{2}{X^{2}} \sprod{\omega_{m}}{d \phi}$,
are then equivalent to the Euler-Lagrange equations for
the Lagrangian
\be
{\cal L}[\Erpot, \phi] \, = \, 
\frac{| d \Erpot \, + \, 2 \bar{\Lambda} d \Lambda |^2}{X^2} 
\, + \, 4 \, \frac{| d \Lambda |^2}{X} 
\label{M-1}
\ee
where $| d \Lambda |^2 \equiv \sprod{d \Lambda}{d \bar{\Lambda}}$, and
where the Ernst potentials $\Erpot$ and $\Lambda$ are given in terms
of the four potentials $X$, $Y$, $\phi$ and $\psi$:
\be
\Erpot \, = \, - \, \left( X + \phi^{2} + \psi^{2} \right)
\, + \, i \, Y \, , \; \; \; \; 
\Lambda \, = \, - \, \phi \, + \, i \, \psi \, .
\label{M-2}
\ee
(It is worth noticing that there is no need to use Weyl 
coordinates and the $(2\!+\!2)$-split of the spacetime metric at this 
point: The structure of the Ernst system and the sigma-model 
identities associated with it concern the symmetries of the 
{\em target\/} manifold. The formulation requires only the existence of 
the above potentials, which can be introduced whenever spacetime 
admits at least {\em one\/} Killing field. The existence of a 
further Killing field (together with the 
integrability properties) is then responsible for the reduction of 
the equations to a boundary value problem on a fixed, $2$-dimensional
domain.)

The key to the uniqueness theorem for rotating electrovac black hole
spacetimes is the observation that the above Lagrangian (\ref{M-1}) 
describes a nonlinear sigma-model on the coset space 
$SU(1,2) / S(U(1) \times U(2))$ \cite{Maz-82-84b}, \cite{Neug}.
(In the vacuum case one obtains, instead, 
a mapping with target manifold $SU(1,1) / U(1)$. If the 
dimensional reduction is performed with respect to the
stationary Killing field, then the target manifold becomes 
$SU(1,2) / S(U(1) \times U(1,1))$.)
In terms of the hermitian matrix $\Phi$,
\be
\Phi_{a b} \, = \, \eta_{a b} \, + \, 2 \, \bar{v}_a v_b \, ,
\; \; \; \mbox{with} \; \; 
(\/ v_0 \/ , \/ v_1 \/ , \/ v_2 \/ ) \, = \, 
\frac{1}{2 \sqrt{X}} \,
(\, \Erpot-1 \, , \, \Erpot+1 \, , \, 2 \Lambda\, ) \, ,
\label{M-3}
\ee
(and $\eta = \mbox{diag} (-1,+1,+1)$), the Lagrangian 
(\ref{M-1}) becomes
\be
{\cal L}[\Phi] \, = \, \frac{1}{2} \, 
\mbox{\em Tr\/} \, \sprod{J}{J} \, , \; \; \; \; \mbox{with} 
\; \; J \, = \, \Phi^{-1} \, d \Phi \, .
\label{M-4}
\ee

The Mazur identity relates the Laplacian of the relative difference 
$\Psi$ of two hermitian matrices $\Phi_{(1)}$ and $\Phi_{(2)}$
to a quadratic expression in $\triangle J$, where
\be
\Psi \, \equiv \, \Phi_{(2)} \Phi_{(1)}^{-1} \, - \, \idid \, ,
\; \; \; \; 
\triangle J \, \equiv \,  J_{(2)} \, - \, J_{(1)}  \, .
\label{M-5}
\ee
Using the general equation 
$\cod (f \alpha) = - \sprod{df}{\alpha} + f \cod \alpha$
for the coderivative of a product of a function $f$ with a
$1$--form $\alpha$, and the properties
$d\Phi = J^{\dagger} \Phi$, $d \Phi^{-1} = - \Phi^{-1} J^{\dagger}$
and $d \Psi = \Phi_{(2)} \/ (\triangle J) \/ \Phi_{(1)}^{-1}$,
it is not hard to derive the identity
\be
-\cod d \, \mbox{Tr} \; \Psi = - \Trace{
\Phi_{(2)} \, \cod (\triangle J) \, \Phi^{-1}_{(1)} } \, + \, 
\mbox{Tr \/} \sprod{\Phi_{(1)}^{-1} \, \triangle J^{\dagger}}
{\Phi_{(2)} \, \triangle J } \, .
\label{M-6}
\ee
If both $\Phi_{(1)}$ and $\Phi_{(2)}$ are {\em solutions\/} to the
variational equations $\cod J = 0$ for the action
$\int {\cal L}[\Phi] \ast 1$, then $\cod (\triangle J)$
vanishes and the first term on the r.h.s. of the Mazur identity 
(\ref{M-6}) does not contribute. 

Finally taking advantage of the full symmetries of spacetime, 
that is, using the circular metric (\ref{metric-circ}),
the coderivative of a stationary and axisymmetric 
$1$--form $\alpha$ becomes
$\cod \alpha  =  \rho^{-1} \cotil \/ (\rho \/ \alpha)$, 
where quantities furnished with a tilde refer to the
$2$-dimensional Riemannian metric 
$\gatens$ (see eq. (\ref{metric-circ})).
The Lagrangian (\ref{M-4}) now describes a nonlinear sigma-model
on the symmetric space $SU(1,2) / S(U(1) \times U(2))$ 
with Riemannian base manifold $(\Gamma, \gatens)$.
The Ernst equations for the positive, hermitian matrix $\Phi$
become $\cotil \/ (\rho \/ J) = 0$. For two solutions
$\Phi_{(1)}$ and $\Phi_{(2)}$, the Mazur identity (\ref{M-6})
therefore yields
\be
\int_{\partial S} \, \rho \, \tilde{\ast} d \, (\mbox{\em Tr\/} \, \Psi) 
\, = \, \int_{S} \rho \, \mbox{Tr \/}
\sprod{\Phi_{(1)}^{-1} \triangle J^{\dagger}}
{\Phi_{(2)} \triangle J} \; \tilde{\eta} \, ,
\label{M-7}
\ee
where $S \subset \Gamma$. 
The uniqueness theorem for the Kerr-Newman metric is now obtained 
as follows: First, on uses the boundary and regularity
conditions to argue that $d (\mbox{Tr} \/ \Psi)$ vanishes on all parts
of the boundary $\partial S$, that is, on the horizon, the rotation 
axis and in the asymptotic regime. The above formula then
implies that the integral on the r.h.s must vanish as well.
Since the embedding of the symmetric space $G/H$ in $G$ can be 
represented in the form $\Phi = g g^{\dagger}$, we may define
${\cal M} \equiv g_{(1)}^{-1} \/ \triangle J^{\dagger} \/ g_{(2)}$,
which yields
\bdm
\mbox{Tr\/} \, \sprod{\Phi_{(1)}^{-1} \triangle J^{\dagger}}
{\Phi_{(2)} \triangle J} \, = \, 
\mbox{Tr\/} \, \sprod{{\cal M}}{{\cal M}^{\dagger}} \, \geq 0 \, .
\edm
Here we have also used the facts that
$\gatens$ is a Riemannian metric, $\triangle J$ is spacelike 
and $\rho \geq 0$. This shows that the r.h.s. of eq. (\ref{M-7})
can only vanish for $\triangle J = 0$, implying that two solutions
$\Phi_{(1)}$ and $\Phi_{(2)}$ with identical boundary and
regularity conditions are equal.

This concludes the outline of the uniqueness proof for the
Kerr-Newman metric amongst the electrovac
black hole solutions with nondegenerate event horizon and 
stationary and axisymmetric, asymptotically flat 
domain of outer communications. The corresponding result
for the vacuum case was already established in
1975 by Robinson \cite{DCR-75}, using a rather complicated
and ingeniously constructed identity.
As it turned out, the Robinson identity is exactly the
Mazur identity for the coset space
$G/H = SU(1,1) / U(1)$ which is, in fact, the relevant
symmetric space for the {\em vacuum\/} Ernst equations.

\section{An Electrovac Bogomol'nyi Equation}

The static electrovac equations imply that the electric
potential $\phi$ depends only on the norm $V$ 
(gravitational potential) of the Killing field.
Israel was able to draw this
conclusion from Stokes' theorem, using a particular
combination of the Einstein-Maxwell equations \cite{{WI-68}}.
Here we present a systematic approach to divergence
identities for electrovac black hole configurations with
nonrotating horizon. We will not assume that the domain
is static. 
Our method yields a set of relations between the
charges and the values of the potentials on the horizon.
Solving for the latter one can then derive the relation
\be
M^2 \, = \, \left[ \frac{1}{4 \pi} \kappa {\cal A} \right]^2
\, + \, Q^2 \, + \, P^2 \, ,
\label{B-0}
\ee
where $M$, $Q$ and $P$ denote the total mass, the 
electric and the magnetic charge, respectively.

We start by recalling the Einstein-Maxwell equations in the
presence of a stationary Killing field $k$, say, 
with norm $V \equiv - \sprod{k}{k}$ and twist 
$\omega \equiv \frac{1}{2} \ast (k \we dk)$. In terms of the
electric potential, $d\phi = -i_kF$, and the magnetic 
potential, $d\psi = i_k \ast F$, the Maxwell equations become
\be
\cod \left( \frac{d \phi}{V} \right) \, = \, 
 - 2 \, \frac{\sprod{\omega}{d \psi}}{V^2}\ \, ,
\; \; \; \; \;  
\cod \left( \frac{d \psi}{V} \right) \, = \, 
2 \, \frac{\sprod{\omega}{d \phi}}{V^2}\ \, .
\label{B-1}
\ee
In addition, we consider the 
identity $\cod(\omega / V^2) = 0$, which gives an
equation for the coderivative of
$\frac{dU}{V^2}$, where $U$ denotes the twist potential, defined by
$\omega = d U  + \psi d \phi -  \phi d \psi$. Finally,
the coderivative of $\frac{dV}{V}$ (i.e., the $R(k,k)$ equation)
becomes
\be
\cod \left( \frac{dV}{V} \right) \, = \, 4 \, 
\frac{\sprod{\omega}{\omega}}{V^2} \, - \,
2 \, \frac{\sprod{d \phi}{d \phi} + \sprod{d \psi}{d \psi}}{V} \, .
\label{B-2}
\ee
(Note that $\cod = \ast d \ast$ is the coderivative with respect to the
spacetime metric.)
Before we apply Stokes' theorem, it is worthwhile noticing that each of
the above equations can be written in the (current conservation) form
$\cod j = 0$. In fact, using for instance 
$\cod(\phi \frac{\omega}{V^2}) = - \frac{\sprod{d \phi}{\omega}}{V^2}$
and $\cod (\psi \frac{d \psi}{V})$ $=$ 
$- \frac{\sprod{d \psi}{d \psi}}{V}$ $+$ 
$2 \psi \frac{\sprod{\omega}{d \phi}}{V^2}$, eqs. (\ref{B-1}) and 
(\ref{B-2}) are equivalent to
\be
\cod \left( \frac{d \phi}{V} 
\, - \, 2  \psi \, \frac{\omega}{V^2} \right) \, = \, 0 \, ,
\; \; \; \; \; 
\cod \left( \frac{d \psi}{V} 
\, + \, 2  \phi \, \frac{\omega}{V^2} \right) \, = \, 0 
\label{B-3}
\ee
and
\be
\cod \left( \frac{dV}{V} \, + 4 U \, \frac{\omega}{V^2} \, - \, 
2 \phi \, \frac{d \phi}{V} \,- \, 2 \psi \, \frac{d \psi}{V} \right)
\, = \, 0 \, , \; \; \; \; \; 
\cod \left( \frac{\omega}{V^2} \right) \, = \, 0 \, ,
\label{B-4}
\ee
respectively.
We can now integrate these equations over a (not necessarily static)
hypersurface extending from the horizon to $S^2_{\infty}$.
Using Stokes' theorem and the fact that all potentials assume
constant values on $H$ and $S^2_{\infty}$, one obtains
linear combinations of the total mass $M$, the charges $Q$ and $P$ and the 
corresponding horizon quantities 
$M_H \equiv \frac{1}{4 \pi} \kappa {\cal A}$, $Q_H$ and $P_H$:
\be
M,  \, (M_H) = \frac{1}{8 \pi} 
\int_{S^2_{\infty}, \, (H)}
\ast \left( k \we \frac{dV}{V} \right) \, ,
\; \; \; \; 
Q, \, (Q_H) = - \frac{1}{4 \pi} 
\int_{S^2_{\infty}, \, (H)}
\ast \left( k \we \frac{d \phi}{V} \right) \, ,
\label{B-5}
\ee
and similarly for $P$ and $P_H$.
(Here we have used the Komar expressions for the charges and
the identities 
$\ast dk$ $=$ $-\frac{2}{V} (k \we \omega)$ $-$ 
$\ast ( k \we \frac{dV}{V} )$ and
$F$ $=$ $k \we \frac{d \phi}{V}$ $+$ $\ast (k \we \frac{ d \psi}{V})$,
as well as the fact that the boundary integrals over
the additional terms vanish; see \cite{{MH-96-LN}} for details.)
For simplicity, we shall also assume that the NUT charge vanishes,
i.e., that $\int_{S^2_{\infty}} d(\frac{k}{V}) = 0$. This
implies that the $1$-form 
$\frac{\omega}{V^2}$ does not contribute
to the boundary integrals, since
$\ast (k \we \frac{\omega}{V^2})$ $=$ $\frac{1}{2} d (\frac{k}{V})$.
Every equation $\cod j  = 0$ gives now rise to a
relation of the form 
$\int_{\infty} \ast(k \we j) = \int_H \ast(k \we j)$, which 
is evaluated by making the following substitutions:
$(dV/V)_{\infty} \rightarrow 2M$, 
$(dV/V)_{H} \rightarrow 2M_H$,
$(d \phi/V)_{\infty} \rightarrow -Q$, 
$(d \phi/V)_{H} \rightarrow -Q_H$,
$(d \psi/V)_{\infty} \rightarrow -P$, 
$(d \psi/V)_{H} \rightarrow -P_H$ and
$(d \omega/V^2)_{\infty} = (d \omega/V^2)_{H} \rightarrow 0$.
Adopting a gauge where 
$\phi_{\infty} = \psi_{\infty} = U_{\infty} = 0$ and using
$V_{\infty} = 1$ and $V_H = 0$, eqs. (\ref{B-3}) and (\ref{B-4}) 
now yield 
\be
Q \, = \, Q_H \, , \; \; \; \; 
P \, = \, P_H \, , \; \; \; \; 
M \, = \, M_H \, + \, \phi_H Q \, + \, \psi_H P \, ,
\label{B-6}
\ee
where the last expression is, of course, the Smarr formula. 

One may now ask whether there exist additional
combinations of the $1$-forms $\alpha^i$
$\equiv$ 
$(\frac{dV}{V},\/ \frac{\omega}{V^2},\/ \frac{d \phi}{V},\/
\frac{d \psi}{V})$
which -- by virtue of the field equations 
(\ref{B-1}) and (\ref{B-2}) -- can be cast into the form of
local conservation laws.
This is indeed the case. A relatively
obvious possibility is
\be
\cod \left( 
\phi \, \frac{d \psi}{V} \, - \, \psi \, \frac{d \phi}{V} 
\, + \, (\phi^2 + \psi^2) \, \frac{\omega}{V^2}
\right) \, = \, 0 \, ,
\label{B-7}
\ee
which, after integration, gives rise to the formula
\be
Q \, \psi_H \, = \, P \, \phi_H \, .
\label{B-8}
\ee
Together with the Smarr formula, this enables one to
solve for the electromagnetic potentials in terms of the
charges, the total mass and the horizon quantity 
$M_H = \frac{1}{4 \pi} \kappa {\cal A}$,
\be
\phi_H \, = \, Q \; \frac{M - M_H}{Q^2 + P^2} \, , \; \; \; \; \; 
\psi_H \, = \, P \; \frac{M - M_H}{Q^2 + P^2} \, .
\label{B-9}
\ee

In order to find all remaining 
differential identities in a systematic way, 
we consider the ansatz 
$j = \sum_{i = 1}^4 f^i[V,U,\phi,\psi] \cdot \alpha^i$ and require
that $\cod j = 0$, i.e., that 
$\sum [f^i \cod \alpha^i - \sprod{d f^i}{\alpha^i}] = 0$.
Writing 
$df = f,_{V} dV + f,_{U} (\omega - \psi d \phi + \phi d \psi) +
f,_{\phi} d \phi + f,_{\psi} d \psi$, one obtains a set of
(simple) partial differential equations for the four functions $f^i$ 
of the four potentials. Solving these equations gives,
in addition to eq. (\ref{B-7}), three formulae
of the desired form. The first one is
\be
\cod \left( [ ... ] \, \frac{\omega}{V^2} +  U \, \frac{dV}{V} +  
\left[\phi^2 + \psi^2 - V \right]
\left[ \phi \frac{d \psi}{V} - \psi \frac{ d \phi}{V} \right] -  
2 \/U \, \left[ \phi \frac{d \phi}{V} + \psi \frac{ d \psi}{V} \right] \,
\right) = 0 ,
\label{B-10-bis}
\ee
(where 
$[ ... ] = \frac{1}{2}(\phi^2 + \psi^2)^2 + 2U^2 - \frac{1}{2}(V^2 +1)$).
This immediately yields 
$- 2 U_H  (M_H + \phi_H Q + \psi_H P ) +
(\phi_H^2 + \psi_H^2)  (\phi_H P - \psi_H Q)= 0$.
Using the Smarr formula (\ref{B-6}) and the
symmetry property (\ref{B-8}), this becomes $U_H \/ M = 0$ and therefore
\be
U_H \, = \, 0 \, ,
\label{B-12}
\ee
which reflects the fact that we consider configurations with
vanishing NUT charge.
The second solution of the differential equations
for the functions $f^i$ gives rise to the conservation law
\be
\cod \left( [ ... ]
\frac{\omega}{V^2} \, - \, \phi \, \frac{dV}{V} \, + \,
\left[\phi^2 - 3 \psi^2 + V -1 \right] \frac{d \phi}{V} \, + \, 
\left[4 \phi \psi - 2 U \right] \frac{d \psi}{V} 
\right) \, = \, 0 \, ,
\label{B-13-bis}
\ee
(where 
$[...] = 2 \psi (1 + \phi^2 + \psi^2) - 4 \phi U$).
(Replacing $\psi \rightarrow \phi$ and $\phi \rightarrow - \psi$ in the
above equation gives the third solution which does, however, 
not contain new information.)
Evaluating the above current on the boundary 
and using again $Q_H = Q$ and 
$P_H = P$ immediately yields
$2 \phi_{H} M_{H} + (\phi_{H}^{2} - 3 \psi_{H}^{2} - 1) \/ Q + 
(4 \phi_{H} \psi_{H} - 2 U_{H}) \/P = 0$. Using the
expressions (\ref{B-9}) and (\ref{B-12}) for the values of the 
potentials in terms of the charges gives the desired formula
(\ref{B-0}) or, equivalently, 
\be
T_H \, = \, \frac{2}{{\cal A}} \, \sqrt{M^2 - Q^2 - P^2} 
\label{B-14}
\ee
for the Hawking temperature $T_H$.
The reason why there exist nontrivial combinations of the Einstein-Maxwell
equations (\ref{B-1}), (\ref{B-2})  which can be cast into the form 
$\cod j = 0$ lies in the structure of the Ernst equations.
We have already mentioned that -- when formulated with respect to the 
stationary Killing field -- the latter are equivalent to a nonlinear
sigma-model with target manifold $SU(1,2) / S(U(1) \times U(1,1))$.
The field equations are therefore obtained from an 
effective Lagrangian of the form (\ref{M-4}) and assume 
the form $\cod J^{a}_{b} = 0$, where $J^{a}_{b}$ is the matrix
valued current $1$-form $\Phi^{-1} d \Phi$. (The definitions of $\Phi$
and of the Ernst potential differ from the definitions given
in eqs. (\ref{M-2}), (\ref{M-3}), since the latter were formed on
the basis of the axial Killing field: Here we use
$\Erpot = \left( -V + \phi^{2} + \psi^{2} \right) +  i Y$ and
$\Phi_{a b} = \eta_{a b} - 2 \bar{v}_a v_b$,
where the Kinnersley vector is defined in terms of the Ernst potentials
$\Erpot$ and $\Lambda$ as before.)
The additional components of the equations 
$\cod J^{a}_{b} = 0$ are, in fact,
exactly the combinations 
(\ref{B-8}), (\ref{B-10-bis}) and (\ref{B-13-bis})
of the Einstein-Maxwell equations
constructed above. 

\vspace{.5cm}

\noindent
I would like to thank the organizing committee of the
{\em Journ\'ees Relativistes 1996\/} for inviting me to give this
lecture. It is a pleasure to acknowledge interesting discussions 
with P. Chru\'sciel, T. Damour, S. Deser, G. Gibbons, W. Israel 
and N. Straumann on various issues of this talk.


\newpage

\begin{thebibliography}{99}
%
\newcommand{\AJM}{{\em Am.\ J.\ Math.\ }}
\newcommand{\AM} {{\em Ann.\ Math.\ }}
\newcommand{\AMC}{{\em Ann.\ Math.\ Soc.\ Coll.\ (4th ed.)\ }}
\newcommand{\ANY}{{\em Ann.\ N.Y.\ Acad.\ Sci.\ }}
\newcommand{\APL}{{\em Ann.\ Physik \ (Leipzig)\ }} 
\newcommand{\APN}{{\em Ann.\ Phys. \ (NY)\ }}
\newcommand{\AP} {{\em Ann.\ Phys.\ }}
\newcommand{\APK}{{\em Ann.\ Physik.\ }}
\newcommand{\APLf}{{\em Ann.\ Physik.\ (Leipzig,\ 4.f.)\ }}
\newcommand{\ASI}{{\em American\ Scientist\ }}
\newcommand{\AIA}{{\em Ann.\ Inst.\ H.\ Poincar\'e\ A\ }}
\newcommand{\AJ} {{\em Astrophys.\ J.\ }}
\newcommand{\BLM}{{\em Bull.\ London\ Math.\ Soc.\ }}
\newcommand{\CQG}{{\em Class.\ Quantum\ Grav.\ }}
\newcommand{\CMP}{{\em Commun.\ Math.\ Phys.\ }}
\newcommand{\CMPshort}{{\em Comm.\ Math.\ Phys.\ }}
\newcommand{\CPA}{{\em Commun.\ Pure \ Appl. \ Math.\ }}
\newcommand{\GRG}{{\em Gen.\ Rel.\ Grav.\ }}
\newcommand{\HPA}{{\em Helv.\ Phys.\ Acta\ }}
\newcommand{\JET}{{\em JETP\ Lett.\ }}
\newcommand{\JDG}{{\em J.\ Diff.\ Geom.\ }}
\newcommand{\JMP}{{\em J.\ Math.\ Phys.\ }}
\newcommand{\JPM}{{\em J.\ Phys.\ A:\ Math.\ Gen.\ }}
\newcommand{\LMP}{{\em Lett.\ Math.\ Phys.\ }}
\newcommand{\MRA}{{\em Mon.\ Not.\ R.\ Astron.\ Soc.\ }}
\newcommand{\NAT}{{\em Nature\ }}
\newcommand{\NPS}{{\em Nature\ (Phys.\ Sci.)\ }}
\newcommand{\NC} {{\em Nuovo\ Cimento\ }}
\newcommand{\NP} {{\em Nucl.\ Phys.\ }}
\newcommand{\NPB}{{\em Nucl.\ Phys.\ B\ }}
\newcommand{\PL} {{\em Phys.\ Lett.\ }}
\newcommand{\PLA}{{\em Phys.\ Lett.\ A\ }}
\newcommand{\PLB}{{\em Phys.\ Lett.\ B\ }}
\newcommand{\PRL}{{\em Phys.\ Rev.\ Lett.\ }}
\newcommand{\PR} {{\em Phys.\ Rev.\ }}
\newcommand{\PRB}{{\em Phys.\ Rev.\ (Sect.\ B)\ }}
\newcommand{\PRP}{{\em Phys.\ Rep.\ }}
\newcommand{\PRD}{{\em Phys.\ Rev.\ D\ }}
\newcommand{\PAW}{{\em Preuss.\ Akad.\ Wiss.\ Berlin,\ Sitz.ber.\ II\ }}
\newcommand{\PTR}{{\em Phil.\ Trans.\ R.\ Soc.\ (London)\ }}
\newcommand{\SDA}{{\em Sitz.ber.\ Deut.\ Akad.\ Wiss.\ Berlin,\ Kl.\ 
Math.-Phys.\ Tech.\ }} 
\newcommand{\SLA}{{\em Proc.\ R.\ Soc.\ London\ Ser.\ A\ }}
\newcommand{\PNA}{{\em Proc.\ Natl.\ Acad.\ Sci.\ }}
\newcommand{\PKN}{{\em Proc.\ Kon.\ Ned.\ Akad.\ Wet.\ }}
\newcommand{\PIA}{{\em Proc.\ Roy.\ Irish\ Acad.\ }}
\newcommand{\PTP}{{\em Progr.\ Theor.\ Phys.\ (Kyoto)\ }}
\newcommand{\RPP}{{\em Rep.\ Prog.\  Phys.\ }}
\newcommand{\RNC}{{\em Riv.\ Nuovo\ Cimento\ }} 
\newcommand{\SJN}{{\em Sov.\ J.\ Nucl.\ Phys.\ }}
\newcommand{\TSM}{{\em Trans.\ Am.\ Math.\ Soc.\ }}
%
%
\bibitem{PC-94-DG}
Chru\'{s}ciel P.T. (1994), in: 
{\em Differential Geometry and Mathematical Physics},
eds. J. Beem \& K.L. Duggal.
Am. Math. Soc., Providence.
%
\bibitem{MH-96-LN}
Heusler M. (1996),
{\em Black Hole Uniqueness Theorems}.
Cambridge Lecture Notes in Physics,
Cambridge Univ. Press.
%
\bibitem{SW-92-93}
Sudarsky D. \& Wald R.M. (1992), \PRD {\bf 46}: 1453;
                         (1993), \PRD {\bf 47}: R5209
%
\bibitem{G-96-CQG}
Galloway G.J. 
(1996), \CQG {\bf 13}: 1471
%
\bibitem{PTC-lect}) 
Chru\'sciel P.T. (1996), 
{\em Uniqueness of Stationary, Electro-Vacuum Black Holes Revisited},
gr-qc/{\bf 9610010}
%
\bibitem{WI-87-300}
Israel W. (1987), in 
{\em 300 Years of Gravitation},
eds. S.W. Hawking \& W. Israel. 
Cambridge University Press.
%
\bibitem{VG-89-JETP}
Volkov M.S. \& Gal'tsov D.V. 
(1989), \JET {\bf 50}: 346;
K\"unzle H.P. \& Masood-ul-Alam A.K.M. 
(1990), \JMP {\bf 31}: 928;
Bizon P. 
(1990), \PRL {\bf 64}: 2844
%
\bibitem{DHS-91-PLB}
Droz S., Heusler M. \& Straumann N. 
(1991), \PLB {\bf 268}: 371;
Heusler M., Droz S. \& Straumann N.
(1991), \PLB {\bf 271}: 61; 
(1992), \PLB {\bf 285}: 21;
Heusler M., Straumann N. \& Zhou Z-H.
(1993), \HPA {\bf 66}: 614;
Luckock H. \& Moss I. 
(1986), \PLB {\bf 176}: 341
%
\bibitem{LM-93-NPB}
Lavrelashvili G. \& Maison D. 
(1993), \NPB {\bf 410}: 407
%
\bibitem{BFM-92-NPB}
Breitenlohner P., Forg\'acs P. \& Maison D. 
(1992), \NPB {\bf 383}: 357;
Greene B.R., Mathur S.D. \& O'Neill C.M. 
(1993), \PRD {\bf 47}: 2242
%
\bibitem{H-72-CMP} 
Hawking S.W. 
(1972), \CMP {\bf 25}: 152
%
\bibitem{HE-73}
Hawking S.W. \& Ellis G.F.R. (1973),
{\em The Large Scale Structure of Space Time}.
Cambridge Univ. Press.
%
\bibitem{KT-66} 
Kundt W. \& Tr\"umper M. (1966),
\APK {\bf 192}: 414
%
\bibitem{C-69}
Carter B. (1969), \JMP {\bf 10}: 70
%
\bibitem{C-87}
Carter B. (1987),
in: {\em Gravitation in Astrophysics}, 
eds. B. Carter \& J.B. Hartle. 
New York, Plenum.
%
\bibitem{MH-93}
Heusler M. (1993), \CQG {\bf 10}: 791
%
\bibitem{PH-73-75} 
Hajicek P. (1973), \PRD {\bf 7}: 2311;
           (1975), \JMP {\bf 16}: 518
%
\bibitem{PC-94-MS} 
Chru\'{s}ciel P.T. \& Wald R.M. 
(1994), \CMP {\bf 163}: 561
%
\bibitem{L-55}
Lichnerowicz A. (1955),
{\em Th\'{e}ories Relativistes de la Gravitation et 
de l'Elec\-tro\-ma\-gn\'{e}\-ti\-sme}.
Paris, Masson.
%
\bibitem{WI-67}
Israel W. (1967), \PR {\bf 164}: 1776
%
\bibitem{WI-68}
Israel W. (1968), \CMP {\bf 8}: 245
%
\bibitem{MZH-73-74}
M\"uller zum Hagen H, Robinson D.C. \& Seifert H.J. 
(1973), \GRG {\bf 4}: 53;
(1974), \GRG {\bf 5}: 61
%
\bibitem{DCR77} 
Robinson D.C. (1977), \GRG {\bf 8}: 695
%
\bibitem{BM-87}
Bunting G.L. \& Masood-ul-Alam A.K.M. (1987), \GRG {\bf 19}: 147
%
\bibitem{MuA-93}
Masood-ul-Alam A.K.M. (1992), \CQG {\bf 9}: L53 
%
\bibitem{WS-85}
Simon W. (1985), \GRG {\bf 17}: 761
%
\bibitem{SY-79-81}
Schoen R. \& Yau S.-T. (1979), \CMP {\bf 65}: 45;
                       (1981), \CMP {\bf 79}: 231
%
\bibitem{Wi-81}
Witten E. (1981), \CMP {\bf 80}: 381
%
\bibitem{Ernst-68}
Ernst F.J. (1968), \PR {\bf 167}: 1175, 
                   \PR {\bf 168}: 1415
%
\bibitem{BC-71-73a-73c}
Carter B. (1971), \PRL {\bf 26}: 331;
          (1973), \NPS {\bf 238}: 71
%
\bibitem{DCR-75}
Robinson D.C. (1975), \PRL {\bf 34}: 905
%
\bibitem{Maz-82-84b}
Mazur P.O. (1982), \JPM {\bf 15}: 3173; 
           (1984), \GRG {\bf 16}: 211, 
                   \PLA {\bf 100}: 341
%
\bibitem{Bunt-83} 
Bunting G.L. (1983),
{\em Proof of the Uniqueness Conjecture for Black Holes}. 
{\em PhD Thesis}, Univ. of New England, Armidale, N.S.W.
%
\bibitem{SD-76} 
Deser S. (1976), \PLB {\bf 64}: 463

\bibitem{Col-77}
Coleman S. (1977), in: 
{\em New Phenomena in Subnuclear Physics},
ed. A. Zichichi.
New York, Plenum.
%
\bibitem{SD-84}
Deser S. (1984), \CQG {\bf 1}: L1
%
\bibitem{BK-88}
Bartnik R. \& McKinnon J. (1988), \PRL {\bf 61}: 141 
%
\bibitem{MHNS-92}
Heusler M. \& Straumann N. (1992), \CQG {\bf 9}:2177
%
\bibitem{BHS-96}
Brodbeck O., Heusler M. \& Straumann N. 
(1996), \PRD {\bf 53}: 754
%
\bibitem{C-70}
Carter B. (1970), \CMP {\bf 17}: 233
%
\bibitem{LS}
Szabados L.B. 
(1987), \JMP {\bf 28}: 2688
%
\bibitem{DS-NH} 
Sudarsky D.
(1995), \CQG {\bf 12}: 579
%
\bibitem{B-NH}
Bekenstein J.D.
(1995), \PRD {\bf 51}: R6608 
%
\bibitem{H-MB}
Heusler M.
(1995), \CQG {\bf 12}: 779
%
\bibitem{P-73}
Penrose R. (1973), \ANY {\bf 224}: 125
%
\bibitem{LV-83}
Ludvigson M \& Vickers J.A.G.
(1983), \JPM {\bf 16}: 3349
%
\bibitem{MM-MB}
Malec E. \& \'O Murchadha N.
(1994), \PRD {\bf 49}: 6931
%
\bibitem{SH-MB}
Hayward S.
(1994), \CQG {\bf 11}: 3037
%
\bibitem{BG-73}
Geroch R.P. (1973), in:
{\em Sixth Texas Symposium on Relativistic Astrophysics},
\ANY {\bf 224}: 108
%
\bibitem{JW-77}
Jang P.S. \& Wald R.M. (1977),
\JMP {\bf 18}: 41;
Jang P.S. (1978), \JMP {\bf 19}: 1152
%
\bibitem{G-PM}
Gibbons G.W., Hawking S.W., Horowitz G.T. \& Perry M.J. (1983),
\CMP {\bf 88}: 295
%
\bibitem{JB-DI}
Bekenstein J.D. (1972), \PRD {\bf 5}: 1239
%
\bibitem{GG}
Gibbons G.W. (1991), in:
{\em The Physical Universe: The Interface Between
Cosmology, Astrophysics and Particle Physics},
Lecture Notes in Physics {\bf 383},
ed. J.D. Barrow.
New York, Springer.
%
\bibitem{FM}
Forg\'acs P. \& Manton N.S. (1980),
\CMP {\bf 72}: 15
%
\bibitem{BH-97}
Brodbeck O. \& Heusler M. (1996),
to appear.
%
\bibitem{MA-RU}
Manton N.S. (1987), \CMP {\bf 111}: 469
%
\bibitem{Ru}
Ruback P. (1988), \CQG {\bf 5}: L155
%
\bibitem{GW}
Weinstein G. (1990), \CPA {\bf 45}: 1183
%
\bibitem{H-ROT}
Heusler M. (1995), \CQG {\bf 12}: 2021
%
\bibitem{JB-BH}
Bekenstein J.D. (1975), \APN {\bf 91}: 75
%
\bibitem{Neug}
Neugebauer G. \& Kramer D. (1969),
\APL {\bf 24}: 62
%
\end{thebibliography}
\end{document}